\documentclass[11pt,final]{iopart_sans}

\usepackage{cite}
\usepackage{graphicx}

\newcommand{\ket}[1]{\vert #1\rangle}
\newcommand{\bra}[1]{\langle #1\vert}
\newcommand{\an}[1]{\hat{#1}}
\newcommand{\cre}[1]{\hat{#1}^\dag}
\newcommand{\cla}[1]{\mathcal{#1}}
\newcommand{\dt}[1]{\frac{\partial #1}{\partial t}}

\newcommand{\ee}{\textrm{e}}
\newcommand{\ri}{\mathrm{i}}
\newcommand{\dd}{\mathrm{d}}
\newcommand{\qq}{\mathbf{q}}
\newcommand{\kk}{\mathbf{k}}
\newcommand{\rr}{\mathbf{r}}

\newcommand{\LL}{\mathcal{L}}
\newcommand{\PP}{\mathcal{P}}
\newcommand{\QQ}{\mathcal{Q}}
\newcommand{\DD}{\mathcal{D}}

\begin{document}

\title{\textsf{Transport of strong-coupling polarons in optical lattices}}

\vspace{0.5cm}

\author{\textsf{M Bruderer}$^{1}$, \textsf{A Klein}$^{1,2}$,
\textsf{S R Clark}$^{1}$ \textsf{and D Jaksch}$^{1,2}$}

\address{$^1$Clarendon Laboratory, University of Oxford,
Parks Road, Oxford OX1 3PU, United Kingdom \\
$^2$ Keble College, Parks Road, Oxford OX1 3PG, United Kingdom}

\ead{m.bruderer@physics.ox.ac.uk}



\begin{abstract}
We study the transport of ultracold impurity atoms immersed in a
Bose--Einstein condensate (BEC) and trapped in a tight optical
lattice. Within the strong-coupling regime, we derive an extended
Hubbard model describing the dynamics of the impurities in terms
of polarons, i.e.~impurities dressed by a coherent state of
Bogoliubov phonons. Using a generalized master equation based on
this microscopic model we show that inelastic and dissipative
phonon scattering results in (i) a crossover from coherent to
incoherent transport of impurities with increasing BEC temperature
and (ii) the emergence of a net atomic current across a tilted
optical lattice. The dependence of the atomic current on the
lattice tilt changes from ohmic conductance to negative
differential conductance within an experimentally accessible
parameter regime. This transition is accurately described by an
Esaki--Tsu-type relation with the effective relaxation time of the
impurities as a temperature-dependent parameter.
\end{abstract}

\date{\today}


\maketitle

\newpage

\section{Introduction}

The study of impurities immersed in liquid helium in the 1960s
opened a new chapter in the understanding of the structure and
dynamics of a Bose-condensed fluid~\cite{Brewer}. Two of the most
prominent examples include dilute solutions of ${}^3$He in
superfluid ${}^4$He~\cite{Brewer,helium-mixtures} and the
measurement of ionic mobilities in superfluid
${}^4$He~\cite{Brewer,PhysRevA.5.356}. More recently, the
experimental realization of impurities in a Bose--Einstein
condensate (BEC)~\cite{PhysRevLett.85.483,PhysRevA.66.043409} and
the possibility to produce quantum degenerate atomic
mixtures~\cite{PhysRevLett.87.080403,PhysRevLett.88.160401,PhysRevLett.89.190404,silber:170408,gunter:180402,ospelkaus:180403}
have generated renewed interest in the physics of impurities. Of
particular importance, collisionally induced transport of
impurities in an ultra-cold bosonic bath has been recently
observed~\cite{ott:160601} in an experimental setup of a similar
type as to the one considered in this paper.

In the context of liquid helium and Bose--Einstein condensates, a
plethora of theoretical results have been obtained. Notably, the
effective interaction between
impurities~\cite{Bardeen-PR-1967,PhysRevA.61.053601,recati:023616,alex-delta-paper}
and the effective mass of
impurities~\cite{PhysRev.94.262,girardeau:279,Gross-motion,PhysRev.127.1452,astrakharchik:013608},
both of which are induced by the condensate background, have been
studied in detail. Moreover, the problem of self-trapping of
static impurities has been
addressed~\cite{PhysRevB.46.301,cucchietti:210401,kalas:043608} in
the framework of Gross-Pitaevskii (GP)
theory~\cite{gross:195,pitaevskii-2003}, and quantized excitations
around the ground state of the self-trapped impurity and the
distorted BEC have been investigated in~\cite{sacha:063604}.

\begin{figure}[h]
\begin{center}
  \includegraphics[width=8cm]{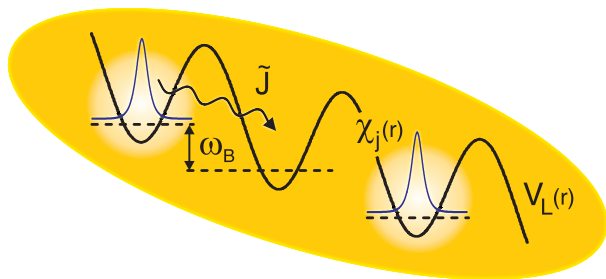}
\end{center}
  \caption{\label{scheme}Impurities described by the localized Wannier functions
  $\chi_j(\rr)$ are trapped in the optical lattice potential $V_L(\rr)$
  and immersed in a nearly uniform BEC. A tilt imposed on the
  periodic potential leads to the formation of a Wannier-Stark
  ladder (dotted lines) with levels separated by the Bloch frequency $\omega_B$.
  Inelastic scattering of Bogoliubov phonons results in a net atomic
  current across the tilted lattice. The current depends on the effective
  hopping $\tilde{J}$, the level separation $\omega_B$ and the
  effective relaxation time of the impurities.}
\end{figure}

In the present paper we study the dynamics of impurities immersed in
a nearly uniform BEC, which in addition are trapped in an optical
lattice potential~\cite{toolbox,bloch-2005}. This setup has been
considered in~\cite{bruderer:011605} by the present authors,
however, we here motivate and develop the small polaron formalism in
more detail and extend the results to the case of a tilted optical
lattice potential, as illustrated in Fig.~\ref{scheme}. In contrast
to similar previously-studied systems, the presence of an optical
lattice allows us to completely control the kinetic energy of the
impurities. Crucially, for impurities in the lowest Bloch band of
the optical lattice the kinetic energy is considerably reduced by
increasing the depth of the lattice
potential~\cite{toolbox,bloch-2005}. As a consequence, it is
possible to access the so-called strong-coupling
regime~\cite{PhysRevB.33.4526}, where the interaction energy due to
the coupling between the impurity and the BEC is much larger than
the kinetic energy of the impurity. This regime is particularly
interesting since it involves incoherent and dissipative
multi-phonon processes which have a profound effect on the transport
properties of the impurities.

As a starting point we investigate the problem of static impurities
based on the GP~approximation and by means of a Bogoliubov
description of the quantized
excitations~\cite{pitaevskii-2003,fetter-AP-1972}. The main part of
the paper is based on small polaron
theory~\cite{Holstein-Ann-1959,mott-alexandrov-1995,Mahan-2000},
where the polaron is composed of an impurity dressed by a coherent
state of Bogoliubov phonons. This formalism accounts in a natural
way for the effects of the BEC background on the impurities and
allows us to describe the BEC deformation around the impurity in
terms of displacement operators. As a principal result, we obtain an
extended Hubbard model~\cite{RevModPhys.62.113,Lewenstein-2007} for
the impurities, which includes the hopping of polarons and the
effective impurity-impurity interaction mediated by the BEC.
Moreover, the microscopic model is particularly suitable for the
investigation of the transport of impurities from first principles
within the framework of a generalized master equation (GME)
approach~\cite{peier,Kenkre-PRB-1975,kenkre-reineker,zwanzig-2001}.

The transport properties of the impurities are considered in two
distinct setups. We first show that in a non-tilted optical
lattice the Bogoliubov phonons induce a crossover from coherent to
incoherent hopping as the BEC temperature
increases~\cite{bruderer:011605}. In addition, we extend the
results in~\cite{bruderer:011605} to the case of a tilted optical
lattice (see Fig.~\ref{scheme}). We demonstrate that the inelastic
phonon scattering responsible for the incoherent hopping of the
impurities also provides the necessary relaxation process required
for the emergence of a net atomic current across the
lattice~\cite{ponomarev:050404}. In particular, we find that the
dependence of the current on the lattice tilt changes from ohmic
conductance to negative differential conductance (NDC) for
sufficiently low BEC temperatures. So far, to our knowledge, NDC
has only been observed in a non-degenerate mixture of ultra-cold
${}^{40}$K and ${}^{87}$Rb atoms~\cite{ott:160601} and in
semiconductor
superlattices~\cite{PhysRevLett.64.3167,PhysRevLett.81.3495},
where the voltage-current dependence is known to obey the
Esaki--Tsu relation~\cite{Esaki-Tsu}. By exploiting the analogy
with a solid state system we show that the transition from ohmic
conductance to NDC is accurately described by a similar relation,
which allows us to determine the effective relaxation time of the
impurities as a function of the BEC temperature.

The implementation of our setup relies on recent experimental
progress in the production of quantum degenerate atomic
mixtures~\cite{PhysRevLett.87.080403,PhysRevLett.88.160401,PhysRevLett.89.190404,silber:170408}
and their confinement in optical lattice
potentials~\cite{gunter:180402,ospelkaus:180403}. As proposed
in~\cite{Griessner-PRL-2006}, the impurities in the optical
lattice can be cooled to extremely low temperatures by exploiting
the collisional interaction with the surrounding BEC. The tilt
imposed on the lattice potential may be achieved either by
accelerating the lattice~\cite{raizen:30} or superimposing an
additional harmonic potential~\cite{ott:160601,fertig:120403}.

An essential requirement for our model is that neither
interactions with impurities nor the trapping potential confining
the impurities impairs the ability of the BEC to sustain
phonon-like excitations. The first condition limits the number of
impurity
atoms~\cite{gunter:180402,ospelkaus:180403,PhysRevB.49.12938}, and
thus we assume that the filling factor of the lattice is much
lower than one, whereas the second requirement can be met by using
a species-specific optical lattice
potential~\cite{leblanc:053612}. Unlike in the case of
self-trapped
impurities~\cite{PhysRevB.46.301,cucchietti:210401,kalas:043608},
we assume that the one-particle states of the impurities are not
modified by the BEC, which can be achieved by sufficiently tight
impurity trapping as shown in \ref{aselftrap}.

The paper is organized as follows. In Section~2 we present our
model. In Section~3, we motivate the small polaron formalism by
considering the problem of static impurities based on the
GP~equation and within the framework of Bogoliubov theory. In
Section~4, the full dynamics of the impurities is included. We
develop the small polaron formalism in detail and derive an
effective Hamiltonian for hopping impurities. In Section~5, the GME
is used to investigate the transport properties of the impurity
atoms. First, we discuss the crossover from coherent to incoherent
hopping and subsequently study the dependence of a net atomic
current across a tilted optical lattice on the system parameters. We
discuss possible extensions of our model and conclude in Section~6.
Throughout the paper we consider impurity hopping only along one
single direction for notational convenience. However, the
generalization to hopping along more than one direction is
straightforward.

\section{Model}

The Hamiltonian of the system is composed of three parts $\hat{H}
= \hat{H}_B + \hat{H}_\chi + \hat{H}_I$, where $\hat{H}_B$ is the
Hamiltonian of the Bose-gas, $\hat{H}_\chi$ governs the dynamics
of the impurity atoms and $\hat{H}_I$ describes the interactions
between the impurities and the condensate atoms.

The Bose-gas is assumed to be at a temperature $T$ well below the
critical temperature $T_c$, so that most of the atoms are in the
Bose-condensed state. The boson-boson interaction is represented
by a pseudo-potential $g\delta(\rr-\rr^\prime)$, where the
coupling constant $g>0$ depends on the s-wave scattering length of
the condensate atoms. Accordingly, the grand canonical Hamiltonian
$\hat{H}_B$ of the Bose-gas reads
\begin{equation}\label{hbec}
   \hat{H}_B = \int\dd\rr\,\cre{\psi}(\rr)\left[\hat{H}_0-\mu_b\right]\an{\psi}(\rr)+
    \frac{g}{2}\int\dd\rr\,\cre{\psi}(\rr)\cre{\psi}(\rr)\an{\psi}(\rr)\an{\psi}(\rr)\,,
\end{equation}
where $\mu_b$ is the chemical potential and
\begin{equation}\label{hbare}
    \hat{H}_0 = -\frac{\hbar^2}{2 m_b}\nabla^2 + V_{\mathrm{ext}}(\rr)
\end{equation}
is the single-particle Hamiltonian of the non-interacting gas, with
$m_b$ the mass of a boson and $V_{\mathrm{ext}}(\rr)$ is a shallow
external trapping potential. The boson field operators
$\cre{\psi}(\rr)$ and $\an{\psi}(\rr)$ create and annihilate an atom
at position $\rr$, respectively, and satisfy the bosonic commutation
relations $[\an{\psi}(\rr),\cre{\psi}(\rr^\prime)] =
\delta(\rr-\rr^\prime)$ and $[\an{\psi}(\rr),\an{\psi}(\rr^\prime)]=
[\cre{\psi}(\rr),\cre{\psi}(\rr^\prime)]=0$.


To be specific we consider \emph{bosonic} impurity atoms with mass
$m_a$ loaded into an optical lattice, which are described by the
impurity field operator $\an{\chi}(\rr)$. The optical lattice
potential for the impurities is of the form $V_L(\rr) = V_x
\sin^2(kx) + V_y \sin^2(ky) + V_z \sin^2(kz)$, with the
wave-vector $k=2\pi/\lambda$ and $\lambda$ the wavelength of the
laser beams. The depth of the lattice $V_\ell$ ($\ell=x,y,z$) is
determined by the intensity of the corresponding laser beam and
conveniently measured in units of the recoil energy $E_R =
\hbar^2k^2/2m_a$. Since we are interested in the strong-coupling
regime we assume that the depth of the lattice exceeds several
recoil energies. As a consequence, the impurity dynamics is
accurately described in the tight-binding approximation by the
Bose--Hubbard model~\cite{toolbox,bloch-2005}, where the
mode-functions of the impurities $\chi_j(\rr)$ are Wannier
functions of the lowest Bloch band localized at site $j$,
satisfying the normalization condition
$\int\dd\rr|\chi_j(\rr)|^2=1$. Accordingly, the impurity field
operator is expanded as $\an{\chi}(\rr) =
\sum_j\chi_j(\rr)\an{a}_j$ and the Bose--Hubbard Hamiltonian is
given by
\begin{equation}\label{bh}
    \hat{H}_\chi = -J\sum_{\langle i,j\rangle} \cre{a}_i\an{a}_j
    +\frac{1}{2}U\sum_j\hat{n}_j(\hat{n}_j-1)
    +\mu_a\sum_j\hat{n}_j + \hbar\omega_B\sum_j j\,\hat{n}_j\,,
\end{equation}
where $\mu_a$ is the energy offset, $U$ is the on-site interaction
strength, $J$ is the hopping matrix element between adjacent sites
and $\langle i,j\rangle$ denotes the sum over nearest neighbours.
The operators $\cre{a}_j$ ($\an{a}_j$) create (annihilate) an
impurity at lattice site $j$ and satisfy the bosonic commutation
relations $[\an{a}_i,\cre{a}_j]=\delta_{i,j}$ and
$[\an{a}_i,\an{a}_j]=[\cre{a}_i,\cre{a}_j]=0$, and $\hat{n}_j =
\cre{a}_j\an{a}_j$ is the number operator. The last term in
Eq.~(\ref{bh}) was added to allow for a tilted lattice with the
energy levels of the Wannier states $\chi_j(\rr)$ separated by Bloch
frequency $\omega_B$, as shown in Fig.~(\ref{scheme}).

For the lattice depths considered in this paper it is possible to
expand the potential wells about their minima and approximate the
Wannier functions by harmonic oscillator ground states. The
oscillation frequencies of the wells are $\omega_\ell = 2\sqrt{E_R
V_\ell}/\hbar$ and the spread of the mode-functions $\chi_j(\rr)$ is
given by the harmonic oscillator length $\sigma_\ell =
\sqrt{\hbar/m_a \omega_\ell}\,$. Explicitly, the Gaussian
mode-function in one dimension reads
\begin{equation}\label{chiapp}
    \chi_{j,\sigma}(x)=\big(\pi\sigma^2\big)^{-1/4}\exp\left[-(x-x_j)^2/(2\sigma^2)\right]\,,
\end{equation}
where $x_j = aj$ with $a=\lambda/2$ being the lattice spacing. We
note that the mode-functions $\chi_{j,\sigma}(\rr)$ are highly
localized on a scale short compared to the lattice spacing $a$
since $\sigma_\ell/a=(V_\ell/E_R)^{-1/4}/\pi\ll 1$ for
sufficiently deep lattices. Consistency of the tight-binging model
requires that excitations into the first excited Bloch band due to
Landau-Zener tunneling or on-site interactions are negligible,
i.e.~$\omega_B\ll\omega_\ell$ and $1/2\,U
n_j(n_j-1)\ll\hbar\omega_\ell$, with $n_j$ the occupation numbers.
These inequalities are readily satisfied in practice.

The density-density interaction between the impurities and the
Bose-gas atoms is again described by a pseudo-potential
$\kappa\delta(\rr-\rr^\prime)$, with $\kappa$ the coupling constant,
and the interaction Hamiltonian is of the form
\begin{equation}\label{hint}
    \hat{H}_I = \kappa
    \int\dd\rr\,\cre{\chi}(\rr)\an{\chi}(\rr)\cre{\psi}(\rr)\an{\psi}(\rr)\,.
\end{equation}
We note that the effect of $\hat{H}_B$ and $\hat{H}_I$ dominate
the dynamics of the system in the strong-coupling regime and hence
the kinetic part in $\hat{H}_\chi$ can be treated as a
perturbation.

\section{Static impurities}\label{static}

In a first step we investigate the interaction of static impurities
with the BEC, i.e.~we neglect the hopping term in $\hat{H}_\chi$. We
first derive a mean-field Hamiltonian based on the GP~approximation
and subsequently quantize the small-amplitude oscillations about the
GP~ground state using standard Bogoliubov
theory~\cite{pitaevskii-2003,fetter-AP-1972}.

In the GP~approximation, the bosonic field operator
$\an{\psi}(\rr)$ is replaced by the classical order parameter of
the condensate $\psi(\rr)$. Accordingly, the impurities are
described by the impurity density $\rho_\chi(\rr) =
\langle\cre{\chi}(\rr)\an{\chi}(\rr)\rangle = \sum_j
n_j|\chi_j(\rr)|^2$, where the average is taken over a fixed
product state $\ket{\Upsilon}$ representing a set of static
impurities, i.e.~$\ket{\Upsilon} = \prod_{\{j\}}\cre{a}_j\ket{0}$,
with $\{j\}$ a set of occupied sites and $\ket{0}$ the impurity
vacuum. We linearize the equation for $\psi(\rr)$ by considering
small deviations $\vartheta(\rr)=\psi(\rr) - \psi_0(\rr)$ from the
ground state of the condensate $\psi_0(\rr)$ in absence of
impurities. This approach is valid provided that
$|\vartheta(\rr)|/\psi_0(\rr)\ll 1$ and corresponds to an
expansion of $\hat{H}_B+\hat{H_I}$ to second order in $\kappa$.
Explicitly, by replacing $\an{\psi}(\rr)$ with $\psi_0(\rr) +
\vartheta(\rr)$ in $\hat{H}_B+\hat{H}_I$ we obtain the
GP~Hamiltonian $H_{GP} = H_{\psi_0} + H_\vartheta +
H_{\mathrm{lin}}$ with\numparts
\begin{equation}\fl
\label{hpsi}
    H_{\psi_0}=\int\dd\rr\left\{\psi_0^\ast(\rr)H_0\psi_0(\rr)-\mu_b|\psi_0(\rr)|^2+
    \frac{g}{2}|\psi_0(\rr)|^4\right\}
    +\kappa\int\dd\rr\,\rho_\chi(\rr)|\psi_0(\rr)|^2\,,
\end{equation}
\begin{eqnarray}\fl\nonumber
 \label{htheta}H_\vartheta&=&\int\dd\rr\left\{\vartheta^\ast(\rr)\left[H_0-\mu_b +
    2g|\psi_0(\rr)|^2\right]\vartheta(\rr)\right\}\\\fl
    &&\qquad\quad+\frac{g}{2}\int\dd\rr\left\{\vartheta^\ast(\rr)[\psi_0(\rr)]^2\vartheta^\ast(\rr)
    +\vartheta(\rr)[\psi_0^\ast(\rr)]^2\vartheta(\rr)\right\}\,,
\end{eqnarray}
\begin{equation}
\label{hlin}\fl
H_{\mathrm{lin}}=\kappa\int\dd\rr\,\rho_\chi(\rr)\left[\psi_0(\rr)\vartheta^\ast(\rr)
+ \psi_0^\ast(\rr)\vartheta(\rr)\right]\,.
\end{equation}\endnumparts
In absence of impurities, i.e.~for $\kappa = 0$ and
$\vartheta(\rr)$ identically zero, the stationary condition
$\delta H_{GP}/\delta\psi_0^\ast(\rr) = 0$ implies that
$\psi_0(\rr)$ satisfies the time-independent GP equation
\begin{equation}\label{gppsi}
    \left[H_0 + g|\psi_0(\rr)|^2\right]\psi_0(\rr) =
    \mu_b\psi_0(\rr)\,.
\end{equation}
The deformed ground state of $H_{GP}$ due to the presence of
impurities is found by imposing the stationary condition $\delta
H_{GP}/\delta\vartheta^\ast(\rr)=0$ or equivalently
\begin{equation}\label{gpelin}
 \left[H_0 - \mu_b  + 2g|\psi_0(\rr)|^2\right]\vartheta(\rr) +
 g[\psi_0(\rr)]^2\vartheta^\ast(\rr)+\kappa\rho_\chi(\rr)\psi_0(\rr)=0\,,
\end{equation}
which is the Bogoliubov--de\,Gennes
equation~\cite{pitaevskii-2003,fetter-AP-1972} for the
zero-frequency mode.

\begin{figure}[t]
\begin{center}
  \includegraphics[width=7.5cm]{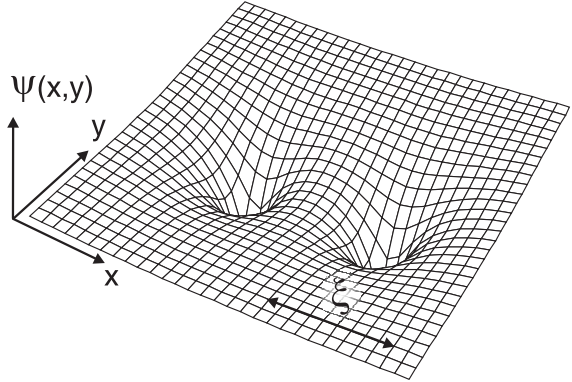}
\end{center}
  \caption{\label{deformation} The order parameter $\psi(\rr)$ of a
  two-dimensional homogeneous condensate deformed by two highly localized
  impurities with $\kappa>0$ according to Eq.~(\ref{solgreen}).
  The deformation of the BEC results in an effective interaction
  potential $\cla{V}_{i,j}$ between the impurities at lattice sites $i$ and $j$.
  The range of the potential is characterized by the healing length $\xi$, which is comparable to
  the lattice spacing $a$ for realistic experimental parameters.}
\end{figure}

For the case of a homogeneous BEC, the effect of the impurities on
the ground state of the condensate can be expressed in terms of
Green's
functions~\cite{PhysRevB.46.301,cucchietti:210401,sacha:063604}.
If we assume for simplicity that both $\psi_0(\rr)$ and
$\vartheta(\rr)$ are real then Eq.~(\ref{gpelin}) reduces to the
modified Helmholtz equation
\begin{equation}\label{gpelinhom}
    \left[\nabla^2 - \left(\frac{2}{\xi}\right)^2\right]\vartheta(\rr) =
    \frac{2\kappa}{g\sqrt{n_0}\xi^D}\rho_\chi(\rr)\,,
\end{equation}
where $\xi=\hbar/\sqrt{m_b g n_0}$ is the healing length, $D$ is
the number of spatial dimensions, and $n_0=|\psi_0|^2$ is the
condensate density. The solution of Eq.~(\ref{gpelinhom}) is found
to be~\cite{arfken-weber}
\begin{equation}\label{solgreen}
    \vartheta(\rr) = -\frac{\kappa}{g\sqrt{n_0}\xi^D}\int
    \dd\rr^\prime\,\mathcal{G}(\rr-\rr^\prime)\rho_\chi(\rr^\prime)\,,
\end{equation}
where the Green's functions are \numparts
\begin{eqnarray}\label{green}
    \mathcal{G}_{\mathrm{1D}}(\rr)&= \frac{1}{2}\exp(-2|\rr|/\xi)\,,\\
    \mathcal{G}_{\mathrm{2D}}(\rr)&= \frac{1}{\pi}\mathrm{K}_0(2|\rr|/\xi)\,,\\
    \mathcal{G}_{\mathrm{3D}}(\rr)&=
    \frac{1}{2\pi}\frac{\exp(-2|\rr|/\xi)}{|\rr|/\xi}\,,
\end{eqnarray}\endnumparts
with $\mathrm{K}_0(x)$ the modified Bessel function of the second
kind. Thus, independent of the dimension $D$ of the system, the
BEC deformation induced by the impurities falls off exponentially
on a length scale set by $\xi$, as illustrated in
Fig.~(\ref{deformation}).

It follows from Eq.~(\ref{solgreen}) that for occupation numbers
$n_j\sim 1$ the condition $|\vartheta(\rr)|/\psi_0(\rr)\ll 1$ is
equivalent to
\begin{equation}\label{condi}
    \alpha=\frac{|\kappa|}{g}\left(\frac{d}{\xi}\right)^D\ll 1\,,
\end{equation}
with $d=n_0^{-1/D}$ the average separation of the condensate
atoms. The generalization of this condition to the non-homogeneous
case is $\alpha(\rr)=(|\kappa|/g)\,[d(\rr)/\xi(\rr)]^D\ll 1$ under
the assumption that the BEC is nearly uniform. We note that
$\alpha(\rr)\propto n_0(\rr)^{(D-2)/2}$, and hence $\alpha(\rr)$
diverges in~1D as $n_0(\rr)\rightarrow 0$, e.g.~near the boundary
of the condensate. However, this is consistent with the GP
approach, which is no longer applicable in the dilute limit
$\xi(\rr)/d(\rr)\ll 1$ of a 1D~Bose-gas~\cite{pitaevskii-2003}.

The change in energy of the BEC due to the impurities is found by
inserting the formal solution for $\vartheta(\rr)$ in
Eq.~(\ref{solgreen}) into $H_{GP}$. Provided that $\vartheta(\rr)$
satisfies Eq.~(\ref{gpelin}) the identity $H_{\mathrm{lin}} +
2H_\vartheta = 0$, or equivalently $H_{GP}=H_{\psi_0} +
\frac{1}{2}H_{\mathrm{lin}}$, holds. Using the latter expression for
$H_{GP}$ we find
\begin{equation}\label{delta}
    H_{GP} = \sum_j\left(\bar{\cla{E}}-\cla{E}_j\right)n_j - \sum_j \cla{E}_j\,n_j\left(n_j-1\right)-\frac{1}{2}\sum_{i\neq j}\cla{V}_{i,j}\,n_i
    n_j\,,
\end{equation}
where $\bar{\cla{E}}= \kappa n_0$ is the first order contribution
to the mean-field shift. In Eq.~(\ref{delta}) we have neglected
constant terms which do not depend on the impurity configuration,
i.e.~contributions containing $\psi_0(\rr)$ only. The off-site
interaction potential between the impurities is given by
\begin{equation}\label{terms}
\cla{V}_{i,j}= \frac{2\kappa^2}{g\xi^D}\int\dd\rr\,\dd\rr^\prime
    |\chi_i(\rr)|^2 \mathcal{G}(\rr-\rr^\prime)|\chi_j(\rr^\prime)|^2\,,
\end{equation}
and $\cla{E}_j = \frac{1}{2}\cla{V}_{j,j}$ is the potential energy
of an impurity, both resulting from the deformation of the
condensate, as illustrated in Fig.~(\ref{deformation}).

In order to quantize the GP solution we consider small excitations
$\an{\zeta}(\rr)$ of the system, which obey bosonic commutation
relations, around the static GP ground state $\psi_0(\rr) +
\vartheta(\rr)$ of the
condensate~\cite{pitaevskii-2003,fetter-AP-1972}. This corresponds
to an expansion of the bosonic field operator as $\an{\psi}(\rr) =
\psi_0(\rr) + \vartheta(\rr) + \an{\zeta}(\rr)$. The Hamiltonian
$\hat{H}_{\zeta}$ governing the evolution of $\an{\zeta}(\rr)$ can
be obtained by substituting $\vartheta(\rr)+\an{\zeta}(\rr)$ for
$\vartheta(\rr)$ in $H_{GP}$. By collecting the terms containing
$\an{\zeta}(\rr)$ and $\cre{\zeta}(\rr)$ we find
\begin{eqnarray}\label{hzeta}\fl\nonumber
 \hat{H}_{\zeta}&=&\int\dd\rr\left\{\cre{\zeta}(\rr)\left[H_0-\mu_b +
    2g|\psi_0(\rr)|^2\right]\an{\zeta}(\rr)\right\}\\\fl
    &&\qquad\quad+\frac{g}{2}\int\dd\rr\left\{\cre{\zeta}(\rr)[\psi_0(\rr)]^2\cre{\zeta}(\rr)
    +\an{\zeta}(\rr)[\psi_0^\ast(\rr)]^2\an{\zeta}(\rr)\right\}\,,
\end{eqnarray}
where the linear terms in $\an{\zeta}(\rr)$ and $\cre{\zeta}(\rr)$
vanish identically since $\vartheta(\rr)$ satisfies
Eq.~(\ref{gpelin}). The Hamiltonian $\hat{H}_{\zeta}$ in
Eq.~(\ref{hzeta}) can be diagonalized by the standard Bogoliubov
transformation \cite{pitaevskii-2003,fetter-AP-1972}
\begin{equation}\label{bog}
    \an{\zeta}(\rr)=\sum_\nu\left[u_\nu(\rr)\an{\beta}_\nu +
    v_\nu^\ast(\rr)\cre{\beta}_\nu\right]\,.
\end{equation}
Here, $u_\nu(\rr)$ and $v_\nu^\ast(\rr)$ are complex functions and
the operators $\cre{\beta}_\nu$ ($\an{\beta}_\nu$) create
(annihilate) a Bogoliubov quasi-particle, with quantum numbers
$\nu$, and satisfy bosonic commutation relations. The Bogoliubov
transformation in Eq.~(\ref{bog}) reduces the Hamiltonian
$\hat{H}_{\zeta}$ to a collection of noninteracting
quasi-particles provided that the coefficients $u_\nu(\rr)$ and
$v_\nu(\rr)$ obey the Bogoliubov--de\,Gennes
equations~\cite{pitaevskii-2003,fetter-AP-1972}
\begin{eqnarray}\label{bogdegen}\eqalign{
    &\left[H_0 - \mu_b  + 2g|\psi_0(\rr)|^2\right] u_\nu(\rr) +
 g[\psi_0(\rr)]^2v_\nu(\rr)= \hbar\omega_\nu u_\nu(\rr)\,,\\
    &\left[H_0 - \mu_b  + 2g|\psi_0(\rr)|^2\right]v_\nu(\rr) +
 g[\psi_0^\ast(\rr)]^2 u_\nu(\rr) = -\hbar\omega_\nu
 v_\nu(\rr)\,,}
\end{eqnarray}
with the spectrum $\hbar\omega_\nu$. Under the condition that
Eqs.~(\ref{bogdegen}) are satisfied one finds that
$\hat{H}_{\zeta}$ takes the form
$\hat{H}_{\zeta}=\sum_\nu\hbar\omega_\nu\cre{\beta}_\nu\an{\beta}_\nu$,
where constant terms depending on $\psi_0(\rr)$ and $v_\nu(\rr)$
only were neglected \footnote{The details of establishing the
above form of $\hat{H}_{\zeta}$ are contained in
Ref.~\cite{fetter-AP-1972}.}.

Consequently, the total Hamiltonian $\hat{H}_\mathrm{stat}$ of the
BEC and the static impurities is composed of three parts
\begin{equation}\label{h2plus}
    \hat{H}_\mathrm{stat} =  \hat{H}_{\zeta} + H_{GP} +  \langle\hat{H}_\chi\rangle\,,
\end{equation}
where $\hat{H}_{\zeta}$ governs the dynamics of the Bogoliubov
quasi-particles and $H_{GP}$ is the GP ground state energy, which
for the case of a homogeneous BEC takes the simple form in
Eq.~(\ref{delta}). The third term $\langle\hat{H}_\chi\rangle$
represents the average value of $\hat{H}_\chi$ with respect to the
fixed product state $\ket{\Upsilon}$ introduced earlier with $n_j$
impurities at each site $j$, giving explicitly
\begin{equation}\label{hav}
    \langle\hat{H}_\chi\rangle = \frac{1}{2}U\sum_j n_j(n_j-1)
    +\mu_a\sum_j n_j + \hbar\omega_B\sum_j j\,n_j\,.
\end{equation}
The ground state of the system corresponds to the Bogoliubov
vacuum defined by $\an{\beta}_\nu\ket{\mathrm{vac}} = 0$.

\section{Hopping impurities in the polaron picture}

Given the Hamiltonian $\hat{H}_\mathrm{stat}$ derived in the
previous section it is straightforward to determine the ground
state energy for a set of static impurities. However, an analysis
of the full dynamics of the impurities requires an alternative
description in terms of polarons, i.e.~impurities dressed by a
coherent state of Bogoliubov quasi-particles. In this picture
small polaron theory allows us, for example, to calculate the
effective hopping matrix element for the impurities, which takes
the BEC background into account.

Our approach is based on the observation that the Hamiltonian
$\hat{H}_{\zeta}$ of the quantized excitations and consequently
the Bogoliubov--de\,Gennes equations~(\ref{bogdegen}) are
independent of $\rho_\chi(\rr)$. In other words, the effect of the
impurities is only to shift the equilibrium position of the modes
$\an{\beta}_\nu$ without changing the spectrum $\hbar\omega_\nu$.
Therefore it is possible to first expand the bosonic field
operator as $\an{\psi}(\rr) = \psi_0(\rr) + \an{\vartheta}(\rr)$,
with $\an{\vartheta}(\rr) = \vartheta(\rr) + \an{\zeta}(\rr)$,
subsequently express $\an{\vartheta}(\rr)$ in terms of Bogoliubov
modes about the state $\psi_0(\rr)$ and finally shift their
equilibrium positions in order to (approximately) minimize the
total energy of the system.

Similarly to the case of static impurities, we first replace
$\an{\psi}(\rr)$ with $\psi_0(\rr) + \an{\vartheta}(\rr)$ in
$\hat{H}_B+\hat{H}_I$ to obtain the Hamiltonian $\hat{H}_{GP} =
\hat{H}_{\psi_0} + \hat{H}_\vartheta + \hat{H}_{\mathrm{lin}}$
with\numparts
\begin{equation}
\label{qhpsi}\fl
    \hat{H}_{\psi_0}=\int\dd\rr\left\{\psi_0^\ast(\rr)H_0\psi_0(\rr)-\mu_b|\psi_0(\rr)|^2+
    \frac{g}{2}|\psi_0(\rr)|^4\right\}
    +\kappa\int\dd\rr\,\cre{\chi}(\rr)\an{\chi}(\rr)|\psi_0(\rr)|^2\,,
\end{equation}
\begin{eqnarray}\fl\nonumber
 \label{qhtheta}\hat{H}_\vartheta&=&\int\dd\rr\left\{\cre{\vartheta}(\rr)\left[H_0-\mu_b +
    2g|\psi_0(\rr)|^2\right]\an{\vartheta}(\rr)\right\}\\\fl
    &&\qquad\quad+\frac{g}{2}\int\dd\rr\left\{\cre{\vartheta}(\rr)[\psi_0(\rr)]^2\cre{\vartheta}(\rr)
    +\an{\vartheta}(\rr)[\psi_0^\ast(\rr)]^2\an{\vartheta}(\rr)\right\}\,,
\end{eqnarray}
\begin{equation}
\label{qhlin}\fl
\hat{H}_{\mathrm{lin}}=\kappa\int\dd\rr\,\cre{\chi}(\rr)\an{\chi}(\rr)\left[\psi_0(\rr)\cre{\vartheta}(\rr)
+ \psi_0^\ast(\rr)\an{\vartheta}(\rr)\right]\,.
\end{equation}\endnumparts
We note that now $\hat{H}_{GP}$ contains the density operator
$\cre{\chi}(\rr)\an{\chi}(\rr)$ instead of $\rho_\chi(\rr)$ in order
to take into account the full dynamics of the impurities. The
expansion of $\an{\vartheta}(\rr)$ in terms of Bogoliubov modes
reads
\begin{equation}\label{thetatrans}
    \an{\vartheta}(\rr) = \sum_\nu\left[u_\nu(\rr)\an{b}_\nu +
    v_\nu^\ast(\rr)\cre{b}_\nu\right]\,,
\end{equation}
where the spectrum $\hbar\omega_\nu$ and coefficients $u_\nu(\rr)$
and $v_\nu(\rr)$ are determined by Eqs.~(\ref{bogdegen}). The
bosonic operators $\cre{b}_\nu$ ($\an{b}_\nu$) create (annihilate)
a Bogoliubov excitation around the ground state of the condensate
$\psi_0(\rr)$ in absence of impurities, and thus, importantly, do
not annihilate the vacuum $\ket{\mathrm{vac}}$ defined in the
previous section, i.e.~$\an{b}_\nu\ket{\mathrm{vac}}\neq 0$. By
substituting the expansion in Eq.~(\ref{thetatrans}) for
$\an{\vartheta}(\rr)$ in the Hamiltonian $\hat{H}_{GP}$ and using
the identity $\cre{\chi}(\rr)\an{\chi}(\rr) =
\sum_{i,j}\chi^\ast_i(\rr)\chi_j(\rr)\cre{a}_i\an{a}_j$ we find
that up to constant terms \footnote{As for $\hat{H}_{\zeta}$, the
details of establishing the above form of $\hat{H}_{\vartheta}$
are contained in Ref.~\cite{fetter-AP-1972}.} \numparts
\begin{eqnarray}\label{qqhpsi}
    \hat{H}_{\psi_0}&=&\sum_{i,j}\bar{\cla{E}}_{i,j}\,\cre{a}_i\an{a}_j\,,\\\label{qqhtheta}
    \hat{H}_\vartheta &=&\sum_\nu\hbar\omega_\nu\cre{b}_\nu\an{b}_\nu\,,\\\label{qqhlin}
    \hat{H}_{\mathrm{lin}}&=&\sum_{i,j,\nu}\hbar\omega_\nu\left[M_{i,j,\nu}\an{b}_\nu
    + M^\ast_{i,j,\nu}\cre{b}_\nu\right]\cre{a}_i\an{a}_j\,,
\end{eqnarray}\endnumparts with the matrix elements
\begin{equation}\label{me0}\eqalign{
    \bar{\cla{E}}_{i,j} &= \kappa\int\dd\rr\,n_0(\rr)\chi^\ast_i(\rr)\chi_j(\rr)\,,\\
    M_{i,j,\nu} &= \frac{\kappa}{\hbar\omega_\nu}\int\dd\rr\,\psi_0(\rr)[u_\nu(\rr)+v_\nu(\rr)]\chi^\ast_i(\rr)\chi_j(\rr)\,,}
\end{equation}
where we assumed for simplicity that $\psi_0(\rr)$ is real. The
non-local couplings $M_{i,j,\nu}$ and $\bar{\cla{E}}_{i,j}$ with
$i\neq j$ resulting from the off-diagonal elements in
$\cre{\chi}(\rr)\an{\chi}(\rr)$ are highly suppressed because the
product of two mode-functions $\chi_i(\rr)$ and $\chi_j(\rr)$ with
$i\neq j$ is exponentially small. As a consequence, the evolution
of the BEC and the impurities, including the full impurity
Hamiltonian $\hat{H}_\chi$, is accurately described by the
Hubbard--Holstein Hamiltonian
\cite{Holstein-Ann-1959,mott-alexandrov-1995,Mahan-2000}
\begin{equation}\label{htot}
    \hat{H}_\mathrm{hol} = \hat{H}_\chi
    +\sum_{j,\nu}\hbar\omega_\nu\left[M_{j,\nu}\an{b}_\nu
    +M^\ast_{j,\nu}\cre{b}_\nu
    \right]\hat{n}_j +\sum_{j}\bar{\cla{E}}_j\,\hat{n}_j
    +\sum_\nu\hbar\omega_\nu\cre{b}_\nu\an{b}_\nu\,,
    \end{equation}
where we discarded the non-local couplings and introduced
$\bar{\cla{E}}_{j} = \bar{\cla{E}}_{j,j}$ and
$M_{j,\nu}=M_{j,j,\nu}$.

Since we intend to treat the dynamics of the impurities as a
perturbation we shift the operators $\an{b}_\nu$ and $\cre{b}_\nu$
in such a way that $\hat{H}_\mathrm{hol}$ in Eq.~(\ref{htot}) is
diagonal, i.e.~the total energy of the system is exactly
minimized, for the case $J=0$. This is achieved by the unitary
Lang-Firsov transformation~\cite{mott-alexandrov-1995,Mahan-2000}
\begin{equation}\label{lftrans}
    \hat{U} =
    \exp\left[\sum_{j,\nu}\left(M^\ast_{j,\nu}\cre{b}_\nu-M_{j,\nu}\an{b}_\nu\right)\hat{n}_j\right]\,,
\end{equation}
from which it follows that by applying the Baker-Campbell-Hausdorff
formula~\cite{barnett-2005} $\hat{U}\cre{b}_\nu\hat{U}^\dag
=\cre{b}_\nu - M_{j,\nu}\hat{n}_j$, $\hat{U}\hat{n}_j\hat{U}^\dag
=\hat{n}_j$ and $\hat{U}\cre{a}_j\hat{U}^\dag =
    \cre{a}_j\cre{X}_j$, with
\begin{equation}\label{polaron}
    \cre{X}_j =
    \exp\left[\sum_\nu\left(M^\ast_{j,\nu}\cre{b}_\nu-M_{j,\nu}\an{b}_\nu\right)\right]\,.
\end{equation}
The operator $\cre{X}_j$ is the sought after displacement operator
that creates a coherent state of Bogoliubov quasi-particles,
i.e.~a condensate deformation around the impurity. As a result,
the transformed Hamiltonian $\hat{H}_{LF}=
\hat{U}\hat{H}_\mathrm{hol}\hat{U}^\dag$ is given by
\begin{eqnarray}\label{hlf}\fl\nonumber
\hat H_{LF}&=&- J \sum_{\langle i,j \rangle}
(\an{X}_i\an{a}_i)^\dag(\an{X}_j\an{a}_j)
+\frac{1}{2}\sum_j\tilde{U}_j\,\hat{n}_j(\hat{n}_j-1)\\\fl
&&\qquad\quad + \sum_j\tilde{\mu}_j\,\hat{n}_j +
\hbar\omega_B\sum_j j\,\hat n_j -\frac{1}{2}\sum_{i \neq j}
V_{i,j} \, \hat n_i \hat n_{j}+
    \sum_\nu\hbar\omega_\nu\cre{b}_\nu\an{b}_\nu\,,
\end{eqnarray}
with the effective energy offset $\tilde{\mu}_j= \mu_a +
\bar{\cla{E}}_j - E_j$, the effective on-site interaction strength
$\tilde{U}_j = U - 2 E_j$, the interaction potential $V_{i,j} =
\sum_\nu\hbar\omega_\nu\left(M_{i,\nu} M^\ast_{j,\nu}
+M^\ast_{i,\nu} M_{j,\nu}\right)$, and $E_j = \frac{1}{2}V_{j,j}$
the so-called polaronic level
shift~\cite{mott-alexandrov-1995,Mahan-2000}.

In the case of static impurities and, more generally, in the limit
$\zeta = J/E_j\ll 1$, the polarons created by $\cre{a}_j\cre{X}_j$
are the appropriate quasi-particles. Consequently, the Hamiltonian
in Eq.~(\ref{hlf}) describes the dynamics of hopping polarons
according to an extended Hubbard
model~\cite{RevModPhys.62.113,Lewenstein-2007} with an
non-retarded interaction potential $V_{i,j}$. The contributions to
$\hat{H}_{LF}$ are qualitatively the same as for static
impurities, except for the additional hopping term. In the
thermodynamic limit, the interaction potential $V_{i,j}$ and the
polaronic level shift $E_j$ are identical, respectively, to their
GP counterparts $\cla{V}_{i,j}$ and $\cla{E}_j$. The reason for
the simple connection between GP theory and small polaron results
is that both involve a linearization of the equations describing
the condensate. It should be noted that the interaction potential
$V_{i,j}$ and the polaronic level shift $E_j$ can also be obtained
exactly from $\hat{H}_\mathrm{hol}$ by applying standard
Rayleigh-Schr\"{o}dinger perturbation theory up to second order in
$\kappa$ since all higher order terms vanish. However, the merit
of using the Lang-Firsov transformation lies in the fact that it
yields a true many-body description of the \emph{state} of the
system, which would require a summation of perturbation terms to
all orders in $\kappa$~\cite{march-1967}.

\begin{figure}[t]
\begin{center}
  \includegraphics[width=5cm]{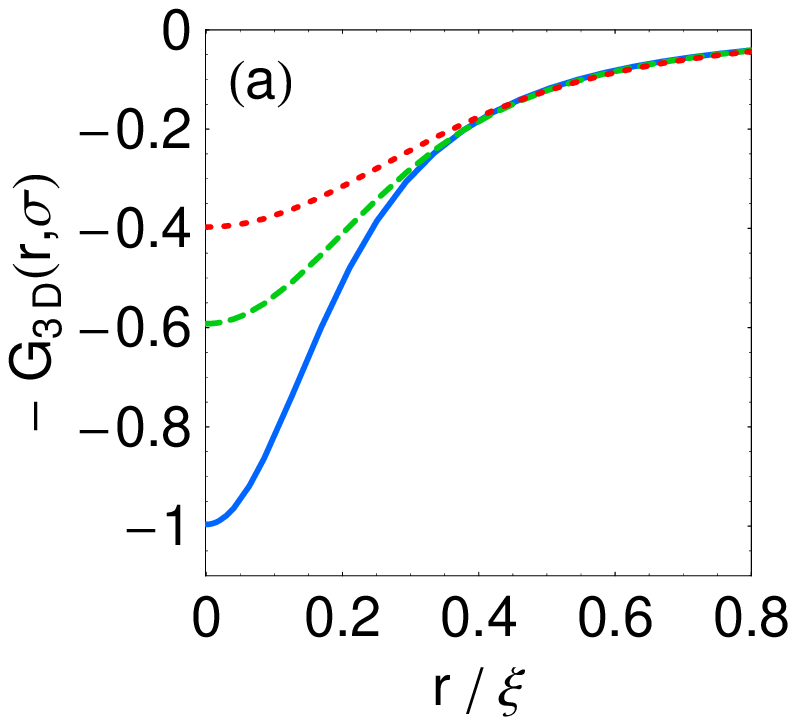}\hspace{0.5cm}\includegraphics[width=5cm]{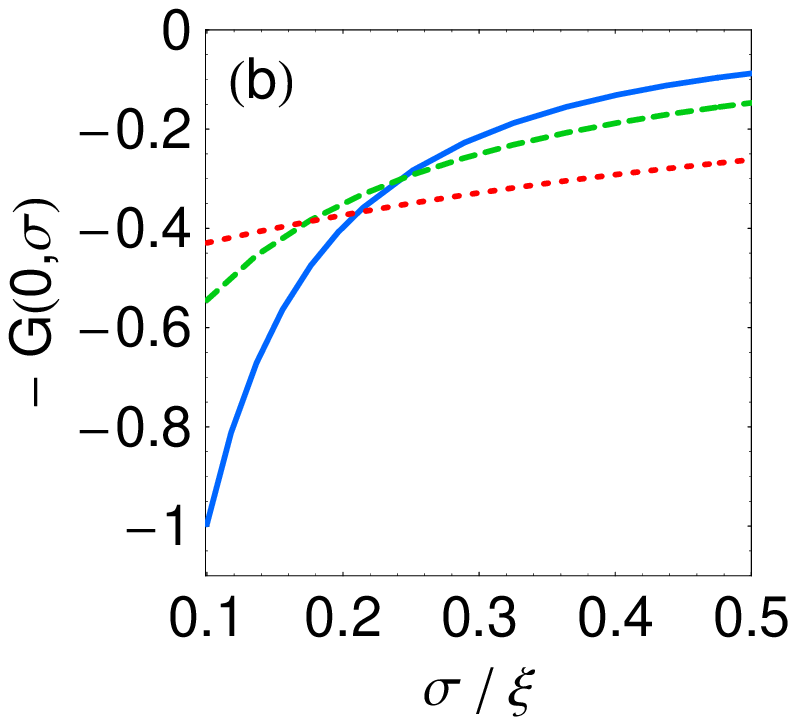}
\end{center}
  \caption{\label{figpot}(a) The function $-\,G_{3D}(\rr,\sigma)$ plotted versus $\rr$ for $\sigma/\xi = 0.1$ (solid line), $\sigma/\xi =
  0.15$ (dashed line), $\sigma/\xi = 0.2$ (dotted line). The potential
  $\cla{V}(\rr)\propto G_{3D}(\rr,\sigma)$ falls off on a scale set by
  $\xi$ and has an increasing depth $\cla{V}(0)$ with decreasing $\sigma$.
  (b) The function $-\,G(0,\sigma)$ plotted versus
  $\sigma$ for 3D (solid line), 2D (dashed line) and 1D (dotted
  line). The depth of the potential $\cla{V}(0)\propto G(0,\sigma)$ and $\cla{E}_p = \frac{1}{2}\cla{V}(0)$ depend strongly on
  $\sigma$ in 3D and only weakly in 2D and 1D. The relevant range of $\sigma$ can be estimated
  for deep lattices by $\sigma/\xi\approx (a/\xi)/[\pi (V_\ell/E_R)^{1/4}]$.}
\end{figure}

We gain qualitative and quantitative insight into the dependence
of the quantities in Eq.~(\ref{hlf}) on the system parameters by
considering the specific case of a homogeneous BEC with the total
Hamiltonian
\begin{eqnarray}\label{hata}\fl\nonumber
\hat H_{LF}&=&- J \sum_{\langle i,j \rangle}
(\an{X}_i\an{a}_i)^\dag(\an{X}_j\an{a}_j)
+\frac{1}{2}\tilde{U}\sum_j\hat{n}_j(\hat{n}_j-1)\\\fl
&&\qquad\quad+\tilde{\mu}\sum_j\hat{n}_j+\hbar\omega_B\sum_jj\,\hat{n}_j-\frac{1}{2}\sum_{i
\neq j} V_{i,j} \, \hat n_i \hat n_{j}+
    \sum_\qq\hbar\omega_\qq\cre{b}_\qq\an{b}_\qq\,,
\end{eqnarray}
with the energy offset $\tilde{\mu}= \mu_a + \kappa n_0 - E_p$,
the on-site interaction strength $\tilde{U} = U - 2 E_p$, the
polaronic level shift $E_j\equiv E_p$ and the phonon momentum
$\qq$. For a homogeneous BEC the Bogoliubov coefficients are of
the form $u_\qq(\rr) = u_\qq\exp(\ri\qq\cdot\rr)$ and $v_\qq(\rr)
= v_\qq\exp(\ri\qq\cdot\rr)$
with~\cite{pitaevskii-2003,fetter-AP-1972}
\begin{equation}\label{bogcoeff}
    u_\qq=\frac{1}{\sqrt{2\Omega}}\bigg(\frac{\varepsilon_\qq +
    gn_0}{\hbar\omega_\qq}+1\bigg)^{1/2}\,\,\mbox{and}\quad
    v_\qq=-\frac{1}{\sqrt{2\Omega}}\bigg(\frac{\varepsilon_\qq +
    gn_0}{\hbar\omega_\qq}-1\bigg)^{1/2}\!.
\end{equation}
Here, $\varepsilon_{\qq} = (\hbar \mathbf{q})^2/2m_b$ is the free
particle energy, $\hbar\omega_\qq= \sqrt{\varepsilon_{\qq}
(\varepsilon_{\mathbf{q}}+2g n_0)}$ is the Bogoliubov dispersion
relation and $\Omega$ is the quantization volume. The corresponding
matrix elements are found to be
\begin{equation}\label{mathom}
    M_{j,\qq} =
    \kappa\sqrt{\frac{n_0\varepsilon_\qq}{\left(\hbar\omega_\qq\right)^3}}f_j(\qq)\,,
\end{equation}
where
\begin{equation}
    f_j(\qq) =
    \frac{1}{\sqrt{\Omega}}\int\dd\rr|\chi_j(\rr)|^2\exp(\ri\qq\cdot\rr)\,.
\end{equation}
For the Gaussian mode-function $\chi_{j,\sigma}(\rr)$ in
Eq.~(\ref{chiapp}) the factor $f_j(\qq)$ takes the simple form
\begin{equation}\label{appf}
    f_j(\qq) =
    \frac{1}{\sqrt{\Omega}}\prod_\ell\exp\left(-q_\ell^2\sigma_\ell^2/4\right)\exp\left(\ri\qq\cdot\rr_j\right)\,.
\end{equation}
The $\qq$-dependence of the impurity-phonon coupling is
$M_{j,\qq}\propto f_j(\qq)/\sqrt{|\qq|}$ in the long wavelength
limit $|\qq|\ll 1/\xi$, which corresponds to a coupling to acoustic
phonons in a solid state system~\cite{Mahan-2000}, whereas in the
free particle regime $|\qq|\gg 1/\xi$ we have $M_{j,\qq}\propto
f_j(\qq)/\qq^2$.

The potential $V_{i,j}$ and the polaronic level shift $E_p$ reduce
to a sum over all momenta~$\qq$ and can be evaluated in the
thermodynamic limit $\Omega^{-1}\sum_\qq\!\rightarrow\!
(2\pi)^{-D}\int\!\dd\qq$ where $V_{i,j}\rightarrow\cla{V}_{i,j}$.
In particular, for the Gaussian mode-functions
$\chi_{j,\sigma}(\rr)$ one finds
\begin{equation}\label{potev}
    \cla{V}_{i,j}  =
    \frac{2\kappa^2}{g\xi^D}\,G(\rr_i-\rr_j,\sigma)\,,
\end{equation}
where the functions $G(\rr,\sigma)$ are defined in \ref{agreen} and
plotted in Fig.~\ref{figpot}a for a three-dimensional system. It
follows from the definition of $G(\rr,0)$ that
$\mathcal{G}(\rr)\equiv G(\rr,0)$, and hence the shape of the
potential is determined by the Green's function $\mathcal{G}(\rr)$
in the limit $\sigma\ll\xi$. The $\sigma$-dependence of
$G(0,\sigma)$, which determines the depth of the interaction
potential $\cla{V}_{i,j}$ and the level shift $\cla{E}_p =
\frac{1}{2}\cla{V}_{j,j}$, is shown in Fig.~\ref{figpot}b. The range
of the potential, characterized by the healing length $\xi$, is
comparable to the lattice spacing $a$ for realistic experimental
parameters, and hence the off-site terms $V_{j,j+1}$ are
non-negligible. In particular, as shown
in~\cite{bruderer:011605,alex-paper}, the off-site interactions can
lead to the aggregation of polarons on adjacent lattice sites into
stable clusters, which are not prone to loss from three-body
inelastic collisions.

\section{Transport}

The coupling to the Bogoliubov phonons via the operators
$\cre{X}_j$ and $\an{X}_j$ changes the transport properties of the
impurities notably. Since we assume the filling factor of the
lattice to be much lower than one we can investigate the transport
properties by considering a single polaron with the Hamiltonian
$\hat{h}_0 + \hat{h}_I$ given by
\begin{eqnarray}\label{ha1}
    \hat{h}_0&= \hbar\omega_B\sum_j j\,\hat{n}_j +
    \sum_\qq\hbar\omega_\qq\cre{b}_\qq\an{b}_\qq\,,\\\label{ha2}
    \hat{h}_I&=-J\sum_{\langle i,j\rangle}
     (\an{X}_i\an{a}_i)^\dag(\an{X}_j\an{a}_j)\,.
\end{eqnarray}
To start we investigate the crossover from coherent to diffusive
hopping~\cite{Holstein-Ann-1959,Mahan-2000} in a non-tilted lattice
($\omega_B=0$) and then extend the result to a tilted lattice
($\omega_B\neq0$) to demonstrate the emergence of a net atomic
current across the lattice~\cite{ponomarev:050404} due to energy
dissipation into the BEC.

\subsection{Coherent versus incoherent transport}

We first consider coherent hopping of polarons at small BEC
temperatures $k_B T\ll E_p$, where incoherent phonon scattering is
highly suppressed. In the strong-coupling regime $\zeta = J/E_j\ll
1$, and hence the hopping term in Eq.~(\ref{ha2}) can be treated
as a perturbation. The degeneracy of the Wannier states requires a
change into the Bloch basis
\begin{equation}\label{blochbasis}
    \ket{\kk} = \frac{1}{\sqrt{N}}\sum_j \exp(\ri
    \kk\cdot\rr_j)\,\cre{a}_j\ket{0}\,,
\end{equation}
with $\kk$ the quasi-momentum and $N$ the number of lattice sites.
Applying standard perturbation theory in the Bloch basis and
describing the state of the system by $\ket{\kk,\{N_\qq\}}$, where
$\{N_\qq\}$ is the phonon configuration with phonon occupation
numbers $N_\qq$, we find the polaron energy up to first order in
$\zeta$
\begin{equation}\label{fop}
    E(\kk)=\tilde{\mu}_a - 2\tilde{J}\cos(\kk\!\cdot\!\mathbf{a})\,,
\end{equation}
where we defined the effective hopping $\tilde{J} =
J\sum_{\qq}\bra{N_\qq}\cre{X}_j\an{X}_{j+1}\ket{N_\qq}$ and
$\mathbf{a}$ is the position vector connecting two nearest
neighbor sites. In particular, for the case of a thermal phonon
distribution with occupation numbers
$N_\qq=(\ee^{\hbar\omega_\qq/k_B T}-1)^{-1}$ the effective hopping
is~\cite{Holstein-Ann-1959,mott-alexandrov-1995,Mahan-2000}
\begin{equation}\label{jtherm}
    \tilde{J} = J\,\exp\Big\{-\sum_{\qq\neq
    0}|M_{0,\qq}|^2\left[1-\cos(\qq\!\cdot\!\mathbf{a})\right]\left(2N_\qq+1\right)\Big\}\,.
\end{equation}
Thus, the hopping bandwidth of the polaron band is highly
suppressed with increasing coupling $\kappa$ and temperature $T$.

At high temperatures $E_p\ll k_B T\ll k_B T_c$ inelastic
scattering, in which phonons are emitted and absorbed, becomes
dominant, and thus the transport of impurities through the lattice
changes from being purely coherent to incoherent. While matrix
elements
$\bra{\{N_\qq\}}\cre{X}_j\an{X}_{j+1}\ket{\{N_\qq\}^\prime}$
involving two different phonon configurations $\{N_\qq\}$ and
$\{N_\qq\}^\prime$ vanish at zero temperature they can take
non-zero values for $T>0$. The condition for energy and momentum
conservation during a hopping event implies that incoherent
hopping is dominated by a three-phonon process that involves
phonons with a linear dispersion $\hbar\omega_\qq = \hbar c|\qq|$,
where $c\sim\sqrt{gn_0/m_b}$ is the speed of sound. This process
is reminiscent of the well-known Beliaev decay of phonons
\cite{pitaevskii-2003}.

\begin{figure}[t]
\begin{center}
\begin{tabular}{ll}
  \raisebox{0.225cm}{\includegraphics[width=4.75cm]{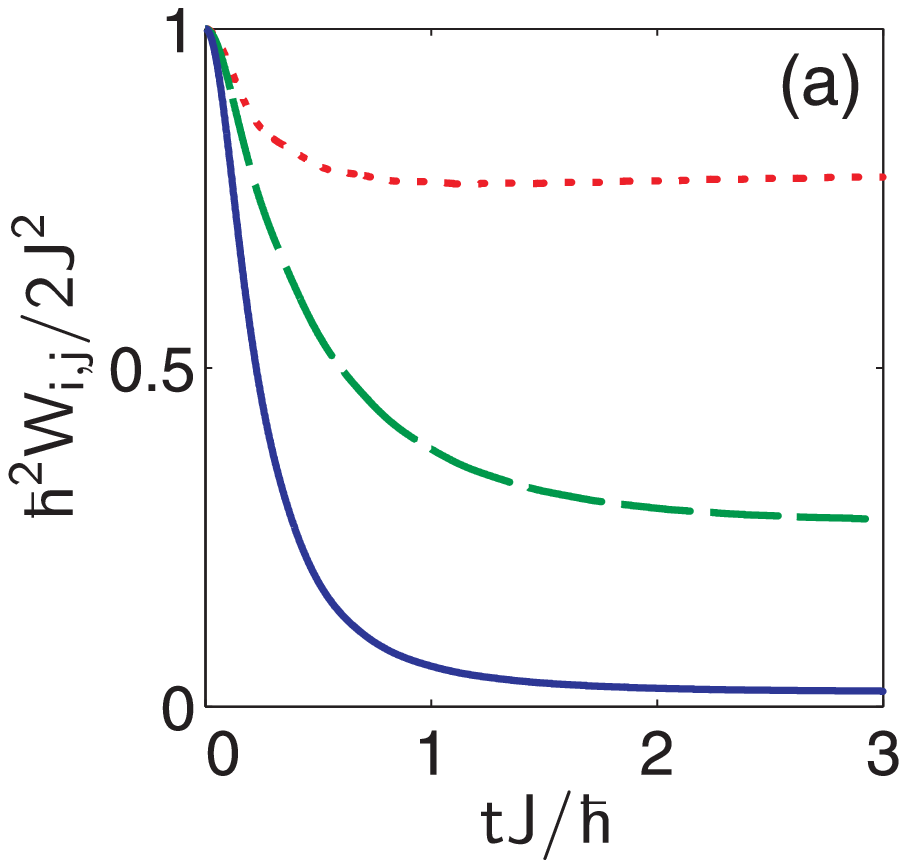}} & \includegraphics[width=5cm]{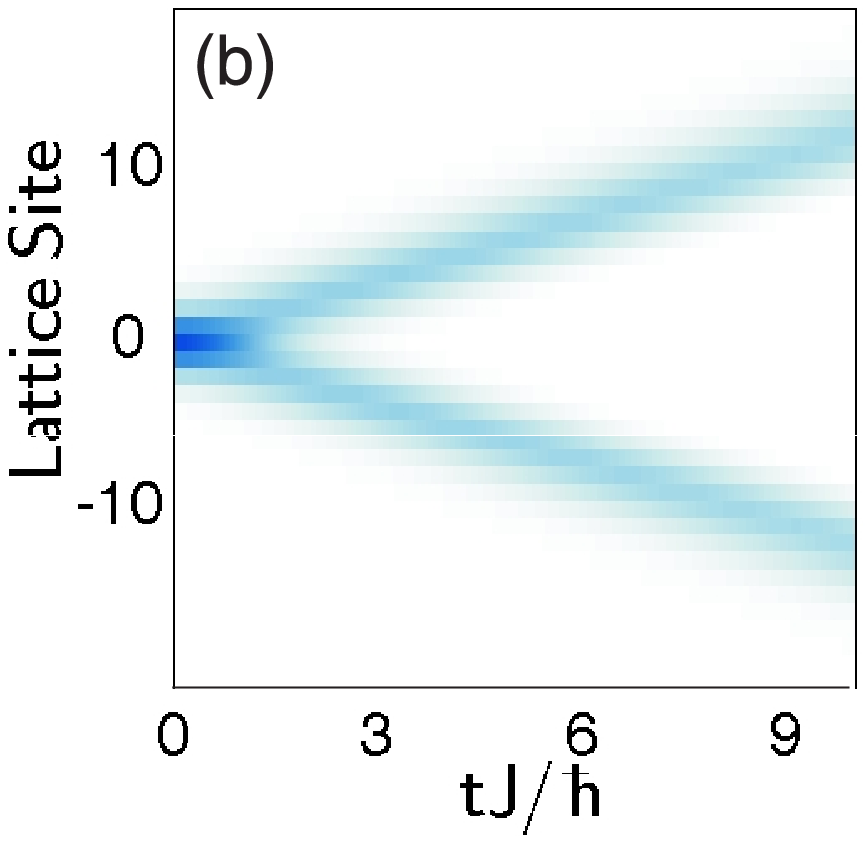} \\
  \includegraphics[width=5cm]{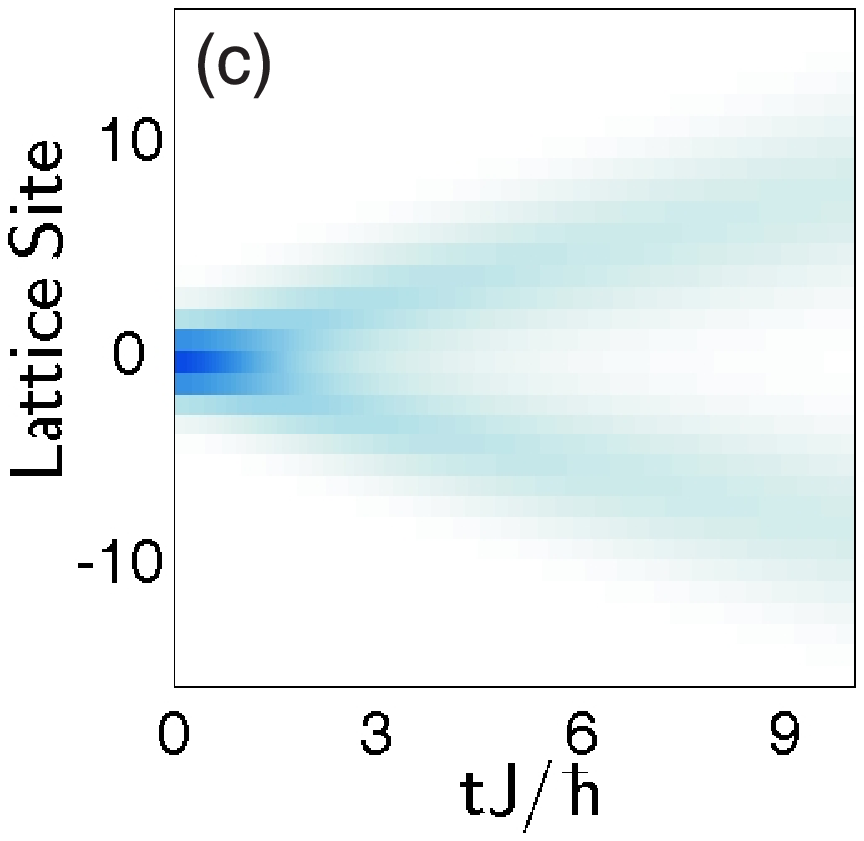} & \includegraphics[width=5cm]{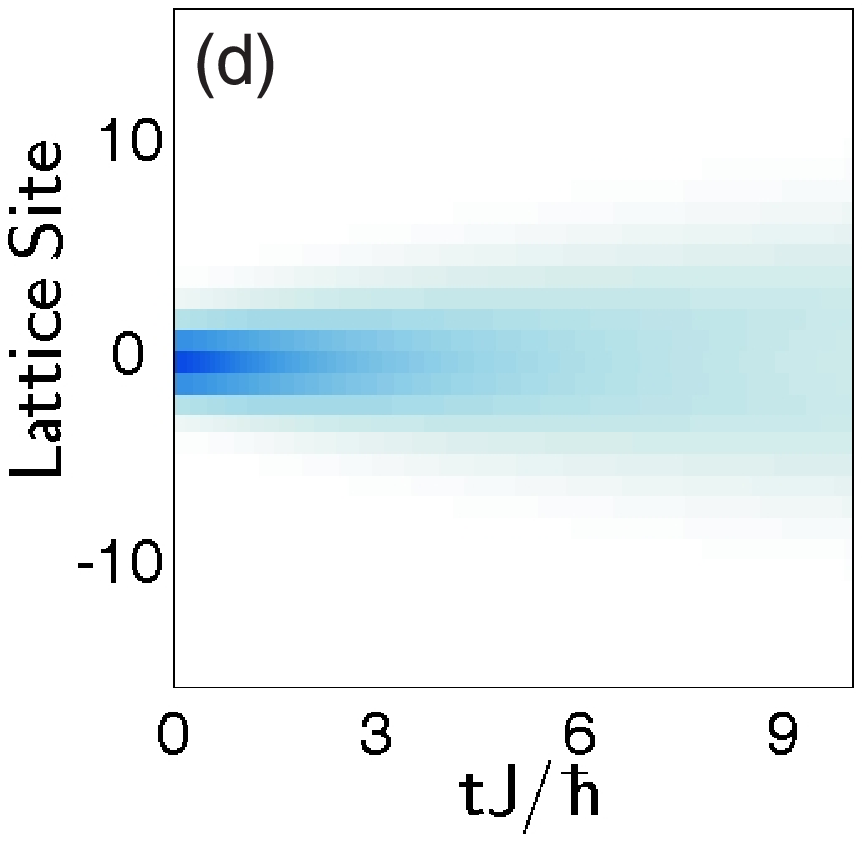} \\
\end{tabular}
\end{center}
\caption{\label{figdiff}Coherent and diffusive hopping in a
one-dimensional system: (a) The memory function
$\hbar^2W_{i,j}(t)/(2J^2)$ plotted versus time for $k_BT = 0$
(dotted line), $k_BT = 5E_p$ (dashed line) and $k_BT = 15E_p$
(solid line). The memory function drops off rapidly for $k_BT\gg
E_p$, indicating the dominance of incoherent hopping. (b) -- (d)
The evolution of the occupation probabilities $P_j(t)$ of a single
impurity initially localized at site $j=0$ according to the GME
with the memory functions in (a). (b) For small temperatures $k_B
T\ll E_p$ the hopping is coherent with two wave-packets moving
away from $j=0$. (d) For high temperatures $k_B T\gg E_p$ the
inelastic scattering of phonons results in a diffusive motion of
the impurity, where the probability $P_j(t)$ remains peaked at
$j=0$. The lattice with spacing $a = 395$nm and $J = 2.45\times
10^{-2}E_R$ contains a single ${}^{41}$K atom. The BEC with $d =
200$nm and $\xi = 652$nm consists of ${}^{87}$Rb atoms;
$\kappa/g=2.58$, $E_p/k_B\approx 10$nK and $J/\hbar\approx
1.2$kHz.}
\end{figure}

We investigate the incoherent transport properties by using a
generalized master equation (GME)~\cite{peier} for the site
occupation probabilities $P_j(t)$ of the impurity. This formalism
has been applied to the transfer of excitons in the presence of
electron-phonon coupling~\cite{Kenkre-PRB-1975,kenkre-reineker}
and is based on the Nakajima-Zwanzig projection
method~\cite{zwanzig-2001}. The generalized master equation is of
the form
\begin{equation}\label{GME}
 \frac{\partial P_i(t)}{\partial t}= \int_0^t \! \dd s \,
 \sum_{j}W_{i,j}(s)\big[P_j(t-s) -  P_i(t-s)\big]\,,
\end{equation}
where the effect of the condensate is encoded in the memory
functions $W_{i,j}(s)$. As shown in \ref{agme} the memory function
to second order in $\zeta$ is given by
\begin{eqnarray}\label{memory}\fl\nonumber
    W_{i,j}(s)&=&\;2\delta_{j,i\pm 1}\Bigg(\frac{J}{\hbar}\Bigg)^2\mathrm{Re}\Bigg[\exp\Big\{\!-2\sum_{\qq \neq 0}
    |M_{0,\qq}|^2
    [1-\cos(\qq\cdot\mathbf{a})] \\\fl
    &&\qquad\times\,[(N_\qq+1)(1-\ee^{\ri\omega_{\qq}s})+
    N_\qq(1-\ee^{-\ri\omega_{\qq}s})]\Big\}\exp(\pm\ri\omega_B s)\Bigg]\,.
\end{eqnarray}
In the case $\omega_B=0$, the nontrivial part of $W_{i,j}(s)$
takes the values $2(J/\hbar)^2$ at $s=0$ and
$2(\tilde{J}/\hbar)^2$ in the limit $s\rightarrow\infty$ due to
the cancellation of highly oscillating terms, as shown in
Fig.~\ref{figdiff}a.

In the regime $k_BT\ll E_p$, the effective hopping $\tilde{J}$ is
comparable to $J$ and the memory function $W_{i,j}$ is well
approximated by $2(\tilde{J}/\hbar)^2\Theta(s)$, with $\Theta(s)$
the Heaviside step function, and thus describes purely coherent
hopping. In the regime $E_p\ll k_BT\ll k_B T_c$, processes involving
thermal phonons become dominant and coherent hopping is highly
suppressed, i.e.~$\tilde{J}\ll J$. In this case the memory function
$W_{i,j}$ drops off sufficiently fast for the Markov approximation
to be valid, as illustrated in Fig.~\ref{figdiff}a. More precisely,
one can replace $P_j(t-s)$ by $P_j(t)$ in Eq.~(\ref{GME}) and after
intergration over $s$ the GME reduces to the standard Pauli master
equation
\begin{equation}\label{pauli}
     \frac{\partial P_i(t)}{\partial t}= \sum_{j}w_{i,j}\big[P_j(t) -  P_i(t)\big]\,,
\end{equation}
where the hopping rates $w_{i,j}$ are given by
\begin{equation}\label{rates}
    w_{i,j} = \int_0^\infty \! \dd s \big[W_{i,j}(s) -
    \lim_{t\rightarrow\infty}W_{i,j}(t)\big]\,.
\end{equation}
The Pauli master equation describes purely incoherent hopping with
a thermally activated hopping rate
$w_{i,j}$~\cite{Holstein-Ann-1959,Mahan-2000}.

The evolution of an initially localized impurity at different
temperatures $T$ for a one-dimensional ${}^{41}$K\,--\,${}^{87}$Rb
system~\cite{PhysRevLett.89.190404} is shown in
Figs.~\ref{figdiff}b~--~\ref{figdiff}d, which were obtained by
numerically solving the GME with the memory function $W_{i,j}(s)$
in Eq.~(\ref{memory}). It can be seen that for small temperatures
$k_B T\ll E_p$ the hopping is coherent with two wave-packets
moving away from the initial position of the impurity atom. In
contrast, for high temperatures $k_B T\gg E_p$ inelastic
scattering of phonons results in a diffusive motion of the
impurity, where the probability $P_j(t)$ remains peaked at the
initial position of the impurity.

\begin{figure}[t]
\begin{center}
\begin{tabular}{ll}
  \includegraphics[height=4.75cm]{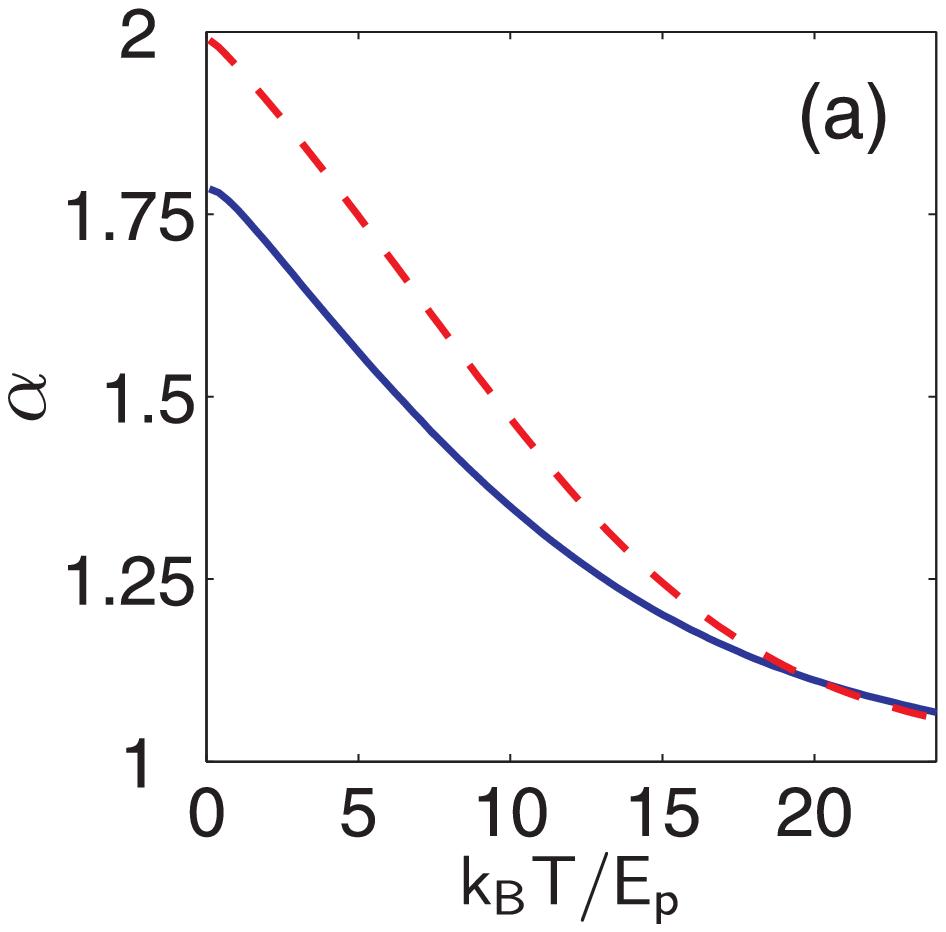}&\includegraphics[height=4.75cm]{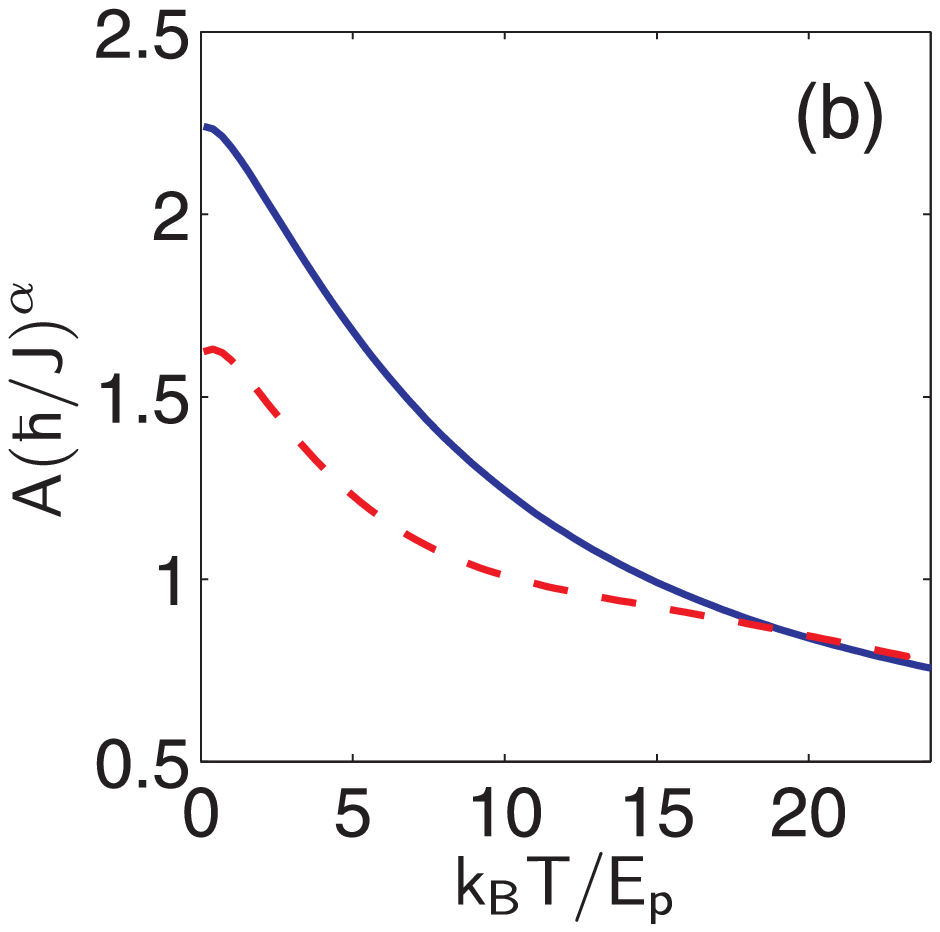} \\
\end{tabular}
\end{center}
  \caption{\label{crossing} Analysis of the mean-square
  displacement $\overline{l^2}(t)=\sum_l l^2 P_l(t)$, obtained
  from the evolution of an initially localized impurity  in a one-dimensional system for the
  time $t_\mathrm{evol}=10\,\hbar/J$ according to the GME. The
  mean-squared displacement was assumed to be of the form
  $\overline{l^2}(t)=A\,t^\alpha$. (a) The exponent $\alpha$
  versus temperature $T$ for the full evolution time (solid line) and the period
  $t_\mathrm{evol}/2$ to $t_\mathrm{evol}$ (dashed line). The drop from $\alpha\approx 2$
  at zero temperature to $\alpha\approx 1$ at high temperatures $k_B T\gg E_p$
  clearly indicates the crossover from coherent to diffusive transport.
  (b) The prefactor $A$  in units of $(J/\hbar)^\alpha$ versus temperature $T$ for the full evolution
  time (solid line) and the period $t_\mathrm{evol}/2$ to $t_\mathrm{evol}$ (dashed line).
  The system parameters are the same as for
  Fig.~\ref{figdiff}.}
\end{figure}

This crossover from coherent to diffusive hopping can be
quantitatively analyzed by considering the mean-squared
displacement of the impurity, $\overline{l^2}(t)=\sum_l l^2
P_l(t)$, which we assume to be of the form
$\overline{l^2}(t)=A\,t^\alpha$. The exponent $\alpha$ takes the
value $\alpha=2$ for a purely coherent process, whereas $\alpha =
1$ for diffusive hopping. Figure~(\ref{crossing}) shows $\alpha$
and $A$ in units of $(J/\hbar)^\alpha$ as functions of the BEC
temperature $T$, which were obtained from the evolution (according
to the GME) of an impurity initially localized at $j=0$. We see
that the exponent $\alpha$ drops from $\alpha\approx 2$ at zero
temperature to $\alpha\approx 1$ at high temperatures $k_B T\gg
E_p$, thereby clearly indicating the crossover from coherent to
diffusive transport. We note that the transition from coherent to
diffusive hopping takes place in a temperature regime accessible
to experimental study and therefore, importantly, may be
observable.

\subsection{Atomic current across a tilted lattice}

The inelastic phonon scattering responsible for the incoherent
hopping of the impurities also provides the necessary relaxation
process required for the emergence of a net atomic current across a
tilted optical lattice. This is in contrast to coherent Bloch
oscillations, which occur in an optical lattice system in absence of
incoherent relaxation effects or dephasing~\cite{raizen:30}. As
pointed out in~\cite{ponomarev:050404}, the dependence of the atomic
current on the lattice tilt $\hbar\omega_B$ changes from ohmic
conductance to NDC in agreement with the theoretical model for
electron transport introduced by Esaki and Tsu~\cite{Esaki-Tsu}.

To demonstrate the emergence of a net atomic current, and, in
particular, to show that impurities exhibit NDC, we consider the
evolution of a localized impurity atom in a one-dimensional system.
With the impurity initially at site $j=0$ we determine its average
position $x_d = \sum_j aj\,P_j(t_d)$ after a fixed drift time $t_d$
of the order of $\hbar/J$. This allows us to determine the drift
velocity $v_d = x_d/t_d$ as a function of the lattice tilt
$\hbar\omega_B$ and the temperature~$T$ of the BEC. In analogy with
a solid state system, the drift velocity $v_d$ and the lattice tilt
$\hbar\omega_B$ correspond to the current and voltage, respectively.

\begin{figure}[t]
\begin{center}
\begin{tabular}{ll}
  \includegraphics[height=5.4cm]{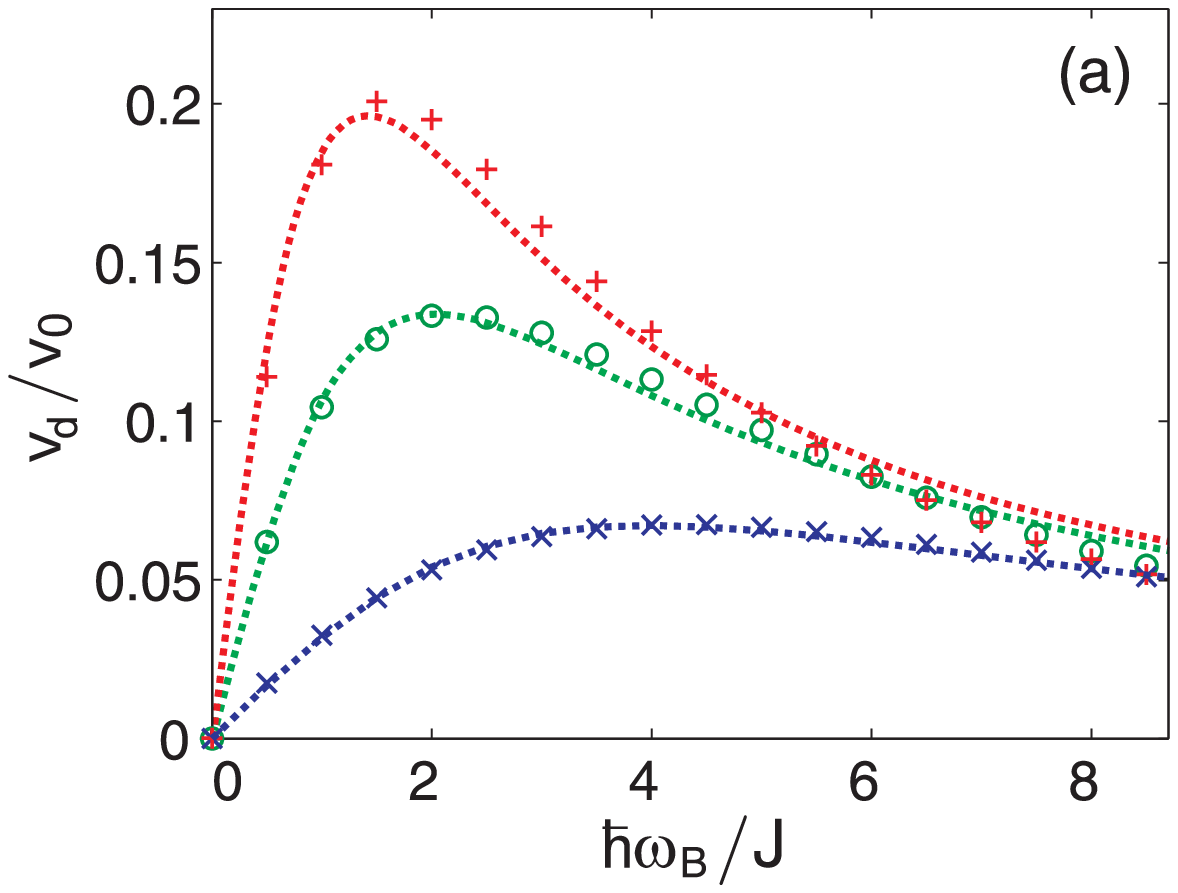}&\raisebox{0.05cm}{\includegraphics[height=5.35cm]{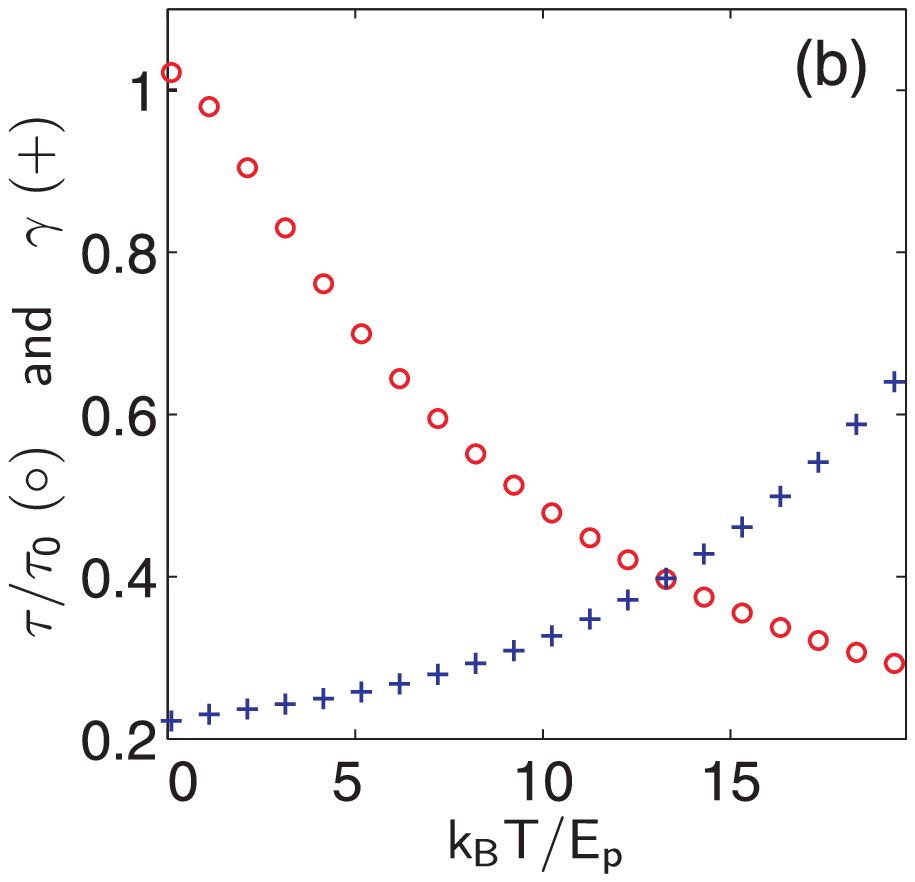}}
\end{tabular}
\end{center}
\vspace{-0.35cm}
  \caption{\label{esaki-tsu}Current-voltage relation for a one-dimensional system:
  (a) The drift velocity $v_d$ in units of $v_0 = Ja/\hbar$ as a function of the
  lattice tilt $\hbar\omega_B/J$ for temperatures $k_BT=0\;(+)$,
  $k_BT=5E_p\;(\circ)$ and $k_BT = 15E_p\;(\times)$ according to the GME and the best fit of the
  Esaki--Tsu-type relation in Eq.~(\ref{esaki}) (dotted lines). The dependence of the current on the tilt
  changes from ohmic conductance $v_d\sim\hbar\omega_B$ to negative differential conductance
  $v_d\sim 1/(\hbar\omega_B)$.
  (b) The relaxation time $\tau\;(\circ)$ in units of $\tau_0 = \hbar/gn_0$ and the prefactor $\gamma\;(+)$
  yielding the best fit of the Esaki--Tsu-type relation to the
  numerical results. The relaxation time $\tau$ of the impurities decreases with increasing BEC temperature $T$ and the
  prefactor $\gamma$ varies only slightly since the exponential temperature dependence of the current is
  accounted for by $\tilde{v}_0$. The system parameters are
  the same as for Fig.~\ref{figdiff}.}
\end{figure}

Figure~\ref{esaki-tsu}a shows the voltage-current relation at
different temperatures $T$, which was obtained by numerically
solving the GME with the memory function in Eq.~(\ref{memory}). We
see that for a small lattice tilt the system exhibits ohmic
behavior $v_d\sim\hbar\omega_B$, whereas for a large lattice tilt
the current decreases with increasing voltage as $v_d\sim
1/(\hbar\omega_B)$, i.e.~the impurities feature NDC.

Following~\cite{ponomarev:050404} we describe the voltage-current
relation for the impurities by an Esaki--Tsu-type relation
\begin{equation}\label{esaki}
    v_d = 2\gamma\tilde{v}_0\,\frac{\omega_B\tau}{1+(\omega_B\tau)^2}\,,
\end{equation}
where  $\tilde{v}_0 = \tilde{J}a/\hbar$ the characteristic drift
velocity, $\tau$ the effective relaxation time of the impurities
and $\gamma$ a dimensionless prefactor. Fitting the
Esaki--Tsu-type relation in Eq.~(\ref{esaki}) to the numerical
data allows us to extract the parameters $\tau$ and $\gamma$, both
depending on the BEC temperature. As can be seen in
Fig.~\ref{esaki-tsu}b, the effective relaxation time $\tau$
decreases with increasing BEC temperature $T$ and is of the order
of $\tau_0  = \hbar/(gn_0)$. The significance of $1/\tau$ is that
of an average collision rate between the impurities and Bogoliubov
excitations, which would allow us, for example, to formulate the
problem of transport in terms of a classical Boltzmann equation
for the distribution function of the impurities~\cite{Mahan-2000}.
The prefactor $\gamma$ varies only slightly since the exponential
dependence of the current on the temperature is accounted for by
$\tilde{v}_0$. We note that independently of $\omega_B\tau$ the
maximum drift velocity is given by $\gamma\,\tilde{v}_0$.

The Esaki--Tsu-type relation in Eq.~(\ref{esaki}) reflects the
competition between coherent and incoherent, dissipative
processes. In the collision dominated regime $\omega_B\tau\ll 1$,
inelastic scattering with phonons destroys Bloch oscillations,
whereas in the collisionless regime $\omega_B\tau\gg 1$ the
evolution of the impurities is mainly coherent, i.e.~Bloch
oscillations of the impurities lead to a suppression of the net
current. The crossover between the two regimes is most pronounced
at zero temperature, where $\tilde{J}$ is comparable to $J$.
However, the change from ohmic to negative differential
conductance is identifiable even at finite temperatures and thus
should be observable in an experimental setup similar to the one
used in~\cite{ott:160601}.

\section{Conclusion}

We have studied the transport of impurity atoms in the
strong-coupling regime, where the interaction energy due to the
coupling between the BEC and the impurities dominates their
dynamics. Within this regime, we have formulated an extended
Hubbard model describing the impurities in terms of polarons,
i.e.~impurities dressed by a coherent state of Bogoliubov phonons.
The model accommodates hopping of polarons and the effective
off-site impurity-impurity interaction mediated by the BEC.

Based on the extended Hubbard model we have shown from first
principles that inelastic phonon scattering results in a crossover
from coherent to incoherent hopping and leads to the emergence of
a net atomic current across a tilted optical lattice. In
particular, we have found that the dependence of the current on
the lattice tilt changes from ohmic conductance to negative
differential conductance for sufficiently low BEC temperatures.
Notably, this transition is accurately described by an
Esaki--Tsu-type relation with the effective relaxation time of the
impurities as a temperature-dependent parameter.

Using the techniques introduced in this paper, qualitatively
similar phenomena can also be shown to occur for fermionic
impurities and, moreover, for impurities of different
species~\cite{taglieber:010401}. For instance, in the case of two
impurity species $A$ and $B$ with the couplings $\kappa_A>0$ and
$\kappa_B<0$, respectively, the effective off-site
impurity-impurity interaction is attractive for the same species,
but repulsive for different species. In either case, observation
of the phenomena reported in this paper lies within the reach of
current experiments, which may give new insight into the interplay
between coherent, incoherent and dissipative processes in
many-body systems.

\ack{MB is indebted to Karl Surmacz for his help with the
manuscript. This work was supported by the UK EPSRC through QIP
IRC (GR/S82176/01) and EuroQUAM project EP/E041612/1, the EU
through the STREP project OLAQUI, the Berrow Scholarship (MB), and
the Keble Association (AK).}

\appendix

\section{Definition of the functions $G(\rr,\sigma)$}\label{agreen}

The function $G(\rr,\sigma)$, which are a generalization of the
Green's functions $\mathcal{G}(\rr)$ for $\sigma>0$, are defined
by
\begin{equation}\label{gg1d}
     G_{\mathrm{1D}}(\rr,\sigma) = \frac{1}{\pi}\int_{-\infty}^\infty\dd
     y\,
     \frac{\cos\left(y|\rr|/\xi\right)}{y^2+4}\exp\left[-\frac{1}{2}\,y^2\left(\frac{\sigma}{\xi}\right)^2\right]\,,
\end{equation}
\begin{equation}\label{gg2d}
     G_{\mathrm{2D}}(\rr,\sigma) = \frac{1}{\pi}\int_0^\infty\dd
     y\,y\,\frac{\mathrm{J}_0\!\left(y|\rr|/\xi\right)}{y^2+4}\exp\left[-\frac{1}{2}\,y^2\left(\frac{\sigma}{\xi}\right)^2\right]\,,
\end{equation}
where $\mathrm{J}_0(x)$ is the Bessel function of the first kind.
\begin{equation}\label{gg3d}
     G_{\mathrm{3D}}(\rr,\sigma) = \frac{1}{\pi^2}\int_0^\infty\dd
     y\,y^2\,\frac{\mathrm{j}_0\!\left(y|\rr|/\xi\right)}{y^2+4}\exp\left[-\frac{1}{2}\,y^2\left(\frac{\sigma}{\xi}\right)^2\right]\,,
\end{equation}
where $\mathrm{j}_0(x) = \sin(x)/x$ is the spherical Bessel
function of the first kind. For the special case $\rr=0$ we have
with $z = \sqrt{2}\,\sigma/\xi$
\begin{eqnarray}\label{gg}
     G_{\mathrm{1D}}(0,\sigma)&=\frac{1}{2}\exp(z^2)\,\mathrm{erfc}(z)\\
     G_{\mathrm{2D}}(0,\sigma)&=-\frac{1}{2\pi}\exp(z^2)\,\mathrm{Ei}(-z^2)\\
     G_{\mathrm{3D}}(0,\sigma)&=\frac{1}{\pi}\left[\frac{1}{\sqrt{\pi}z}- \exp(z^2)\,\mathrm{erfc}(z)\right]
\end{eqnarray}
where $\mathrm{erfc}(x)$ is the complementary error function and
$\mathrm{Ei}(x)$ is the exponential
integral~\cite{abramowitz-stegun}.

\section{Derivation of the GME}\label{agme}

The Hamiltonian for the single impurity and the BEC is
\begin{eqnarray}\nonumber
    \hat{h}_0&=\hbar\omega_B\sum_j j\,\hat{n}_j + \sum_\qq\hbar\omega_\qq\cre{b}_\qq\an{b}_\qq\\\nonumber
    \hat{h}_I&=-J\sum_{\langle i,j\rangle}
     (\an{X}_i\an{a}_i)^\dag(\an{X}_j\an{a}_j)\,,
\end{eqnarray}
where $\hat{h}_0$ is the unperturbed Hamiltonian and $\hat{h}_I$ is
treated as a perturbation in the strong-coupling regime $\zeta\ll
1$.

Starting point of the derivation of the GME is the
Liouville-von\,Neumann equation~\cite{peier}
\begin{equation}\label{lvn}
    \dt{\hat{\rho}(t)} = \LL\hat{\rho}(t)\,,
\end{equation}
where $\hat{\rho}(t)$ is the density matrix of the system
expressed in the eigenbasis of $\hat{h}_0$. The Liouville operator
$\LL$ is defined by $\LL = \LL_0 + \LL_I$, with
$\LL_0=-\ri/\hbar\,[\hat{h}_0,\,\cdot\,]$ and
$\LL_I=-\ri/\hbar\,[\hat{h}_I,\,\cdot\,]$. The density matrix
$\hat{\rho}(t)$ is decomposed as $\hat{\rho}(t) = \PP\hat{\rho}(t)
+ \QQ\hat{\rho}(t)$, where $\PP$ is the projection operator on the
relevant part of $\hat{\rho}(t)$ and $\QQ = 1-\PP$ the
complementary projection operator to $\PP$. The time evolution of
the relevant part $\PP\hat{\rho}(t)$ is governed by the
Nakajima-Zwanzig equation~\cite{zwanzig-2001}
\begin{equation}\label{nze}
    \dt{}\PP\hat{\rho}(t) = \PP\LL\PP\hat{\rho}(t) +\!\int_0^t\!\dd s\,\PP\LL\ee^{\QQ\LL
    s}\QQ\LL\PP\hat{\rho}(t-s) + \PP\LL\ee^{\QQ\LL
    t}\QQ\hat{\rho}(0)\,.
\end{equation}
Specifically, the projection operator $\PP$ employed in the
derivation of the GME is defined by $\PP\hat{\rho}(t) =
\hat{\rho}_B\otimes\DD\Tr_B\hat{\rho}(t)$~\cite{peier,Kenkre-PRB-1975}.
Here, $\hat{\rho}_B$ is the density matrix of the condensate in
thermal equilibrium, $\Tr_B$ is the trace over the condensate
degrees of freedom and $\DD$ is the projection operator on the
diagonal part. Thus $\hat{\varrho}_A(t)=\DD\Tr_B\hat{\rho}(t)$ is
the diagonal part of the reduced density matrix. We note that the
trace over the Bogoliubov phonon states in the definition of $\PP$
introduces irreversibility into the system.

The Nakajima-Zwanzig equation can be simplified given that
$\PP\LL_0 = \LL_0\PP = 0$ and $\PP\LL_I\PP=0$ for the definitions
of $\LL_0$, $\LL_I$ and $\PP$ above. In addition, we assume that
the initial density matrix of the total system has the form
$\hat{\rho}(0)=\hat{\rho}_B\otimes\hat{\varrho}_A(0)$ so that
$\QQ\hat{\rho}(0) = 0$, and hence the inhomogeneous term in
Eq.~(\ref{nze}) vanishes. Taking these simplification into account
and using $\Tr_B\PP\hat{\rho} = \hat{\varrho}_A$ we find from
Eq.~(\ref{nze}) that $\hat{\varrho}_A(t)$ evolves according
to~\cite{peier}
\begin{equation}\label{nze2}
    \dt{}\hat{\varrho}_A(t) =  \int_0^t\dd
    s\,K(s)\hat{\varrho}_A(t-s)
\end{equation}
with the memory kernel
\begin{equation}\label{kernel}
    K(s) = \DD\Tr_B\left[\LL_I\ee^{\LL_0s+\QQ\LL_Is}\LL_I\,\hat{\rho}_B(0)\right]\,.
\end{equation}
The memory kernel to second order in $\zeta$, i.e.~dropping
$\LL_I$ in the exponent in Eq.~(\ref{kernel}), can be expressed in
tetradic form as~\cite{Kenkre-PRB-1975,zwanzig-2001}
\begin{eqnarray}\label{kernel0}\fl\nonumber
  K^0_{ii,jj}(s)&=&\;\frac{2}{\hbar^2}\,Z_B^{-1}\hspace{-10pt}\sum_{\{N_\qq\},\{N_\qq\}^\prime}\hspace{-5pt}\ee^{-\hbar\omega_{\{N_\qq\}}/k_BT}|\bra{i,\{N_\qq\}^\prime}\hat{h}_I\ket{j,\{N_\qq\}}|^2\\\fl
  &&\qquad\times\cos\left[(\Omega_{i,j} + \hbar\omega_{\{N_\qq\}}-\hbar\omega_{\{N_\qq\}^\prime})\,t\right]\,,
\end{eqnarray}
with the partition function of the condensate
\begin{equation}\label{partition}
    Z_B = \sum_{\{N_\qq\}}\ee^{-\hbar\omega_{\{N_\qq\}}/k_BT}\,,
\end{equation}
and where $\hbar\Omega_{i,j}$ is the energy difference between the
impurity configurations $i$ and $j$, and $\hbar\omega_{\{N_\qq\}}$
is the energy of the phonon configuration $\{N_\qq\}$.

The explicit expression for the memory function $W_{i,j}(s) =
K^0_{ii,jj}(s)$ for the Hamiltonian $\hat{h}_0 + \hat{h}_I$ has
been evaluated in~\cite{Kenkre-PRB-1975,kenkre-reineker} based on
Eq.~(\ref{kernel}), however, we here give an alternative
derivation of $W_{i,j}(s)$ starting from Eq.~(\ref{kernel0}). The
evaluation of $K^0_{ii,jj}(s)$ in Eq.~(\ref{kernel0}) can be
separated into a phonon part and an impurity part, where the
latter is given by
\begin{equation}\label{imppar}
    J^2\sum_k|\bra{i}\cre{a}_{k}\an{a}_{k\pm 1}\ket{j}|^2\,\ee^{-\ri\Omega_{i,j}t} =
    J^2\delta_{j,i\pm1}\,\ee^{\pm\ri\omega_B
    t}\,.
\end{equation}
Thus, for the phonon part we only have to consider operators of
the form $\cre{X}_{j}\an{X}_{j\pm 1}$, which can be written in
terms of displacement operators $D(\beta) = \exp(\beta\cre{b} -
\beta^\ast\an{b})$ as
\begin{equation}\label{product}
    \cre{X}_j\an{X}_{j\pm 1} =
    \prod_{\qq}D(\beta_{j,\qq})\,\ee^{\ri\Phi_{j,\qq}}\,,
\end{equation}
with $\beta_{j,\qq} = M^\ast_{j,\qq} - M^\ast_{j\pm 1,\qq}$ and
$\Phi_{j,\qq}$ the corresponding phase. This allows us to treat each
phonon mode in Eq.~(\ref{kernel0}) separately, and the problem
reduces to the summation
\begin{equation}\label{sumsum}
    \sum_{n=0}^\infty\sum_{m=0}^\infty\ee^{-n\hbar\omega_\qq/k_B T}|\bra{m}D(\beta_{j,\qq})\ket{n}|^2\,\ee^{\ri\omega_\qq t (m - n)}\,,
\end{equation}
where $m$ and $n$ are phonon occupation numbers of the mode
$\omega_\qq$. The matrix elements of the displacement operator in
the Fock basis are~\cite{barnett-2005}
\begin{equation}
\begin{array}{lcl}
   \bra{m}D(\beta)\ket{n} = \sqrt{\frac{n!}{m!}}\beta^{m-n}\ee^{-|\beta|^2/2}\mathrm{L}_n^{m-n}(|\beta|^2) & \mbox{for} & m\geq n\vspace{10pt}\\
   \bra{m}D(\beta)\ket{n} = \sqrt{\frac{m!}{n!}}(-\beta^\ast)^{n-m}\ee^{-|\beta|^2/2}\mathrm{L}_m^{n-m}(|\beta|^2) & \mbox{for} & n\geq m\,,
\end{array}
\end{equation}
where $\mathrm{L}_n^k(x)$ are generalized Laguerre polynomials. At
this point we introduce the new variables $x=|\beta_{j,\qq}|^2$,
$y=\ee^{\ri\omega_\qq t}$, $z=\ee^{-\hbar\omega_\qq/k_BT}$, and $l =
m - n$. For the case $m\geq n$, or equivalently $l\geq 0$, we find
after the substitution of $m = n + l$ that the sum in
Eq.~(\ref{sumsum}) becomes
 \begin{equation}\label{mgn}
        \ee^{-x}\sum_{l=0}^\infty(xy)^l\sum_{n=0}^\infty\frac{n!}{(n+l)!}\left[\mathrm{L}_n^l(x)\right]^2z^n\,.
\end{equation}
To evaluate the sum over $n$ we use the fact that the following
relation for generalized Laguerre polynomials
holds~\cite{gradstein}
\begin{equation}\label{rel}
    \sum_{n=0}^\infty
    n!\frac{\mathrm{L}_n^{\gamma}(x)\mathrm{L}_n^{\gamma}(y)z^n}{\Gamma(n+\gamma+1)}=
    \frac{(xyz)^{-\gamma/2}}{1-z}\exp\left(-z\frac{x+y}{1-z}\right)\mathrm{I}_{\gamma}\left(2\frac{\sqrt{xyz}}{1-z}\right)\,,
\end{equation}
provided that $|z|<1$. Here, $\Gamma(x)$ is the Gamma function and
$\mathrm{I}_\gamma(x)$ are modified Bessel functions of the first
kind. Using the relation in Eq.~(\ref{rel}) we find that the sum
over $n$ yields
\begin{equation}
    \frac{1}{1-z}\exp\left(-\frac{2xz}{1-z}\right)(x\sqrt{z})^{-l}\mathrm{I}_l\left(\frac{2x\sqrt{z}}{1-z}\right)\,.
\end{equation}
For the case $n\geq m$, or equivalently $l\leq 0 $, we have to
express $\mathrm{L}_m^{n-m}(x)$ in terms of
$\mathrm{L}_n^{m-n}(x)$ in order to exploit Eq.~(\ref{rel}). Using
the relation
\begin{equation}
    \mathrm{L}_r^{-s}(x) = \frac{x^s}{(-r)_s}\mathrm{L}_{r-s}^s(x)\,,
\end{equation}
with $(x)_n = x(x+1)(x+2)\cdots(x+n-1)$ the Pochhammer symbol, we
find that
\begin{equation}
    \left[\mathrm{L}_m^{-(m-n)}(x)\right]^2=\frac{x^{2(m-n)}}{\left[(-m)_{(m-n)}\right]^2}\left[\mathrm{L}_n^{m-n}(x)\right]^2\,,
\end{equation}
and since $\left[(-m)_{(m-n)}\right]^2 =
\left[(-m)(-m+1)\cdots(-n-1)\right]^2=(m!/n!)^2$ substituting
$l=m-n$ in expression (\ref{sumsum}) yields
\begin{equation}
    \ee^{-x}\sum_{l=0}^{-\infty}
        (xy)^l\sum_{n=0}^\infty\frac{n!}{(n+l)!}\left[\mathrm{L}_n^l(x)\right]^2z^n\,,
\end{equation}
which is identical to expression (\ref{mgn}) except for the upper
limit in the sum over $l$. Using relation (\ref{rel}) again and
discarding the double counting of $l=0$ we find that the total sum
in Eq.~(\ref{sumsum}) is given by
\begin{equation}\label{lsum}
        \frac{1}{1-z}\exp\left[-x\left(1+\frac{2z}{1-z}\right)\right]\sum_{l=-\infty}^{\infty}\left(\frac{y}{\sqrt{z}}\right)^l\mathrm{I}_l\left(\frac{2x\sqrt{z}}{1-z}\right)\,.
\end{equation}
To evaluate the sum over $l$ we use the
identity~\cite{abramowitz-stegun}
\begin{equation}
    \sum_{l=-\infty}^{\infty}
    \mathrm{I}_l(x)t^l=\exp\left[\frac{x}{2}\left(t+\frac{1}{t}\right)\right]
\end{equation}
and find that Eq.~(\ref{lsum}) equals
\begin{equation}
    \frac{1}{1-z}\exp\left[-x\left(1+\frac{2z}{1-z}\right)\right]\exp\left[\frac{x\sqrt{z}}{1-z}\left(\frac{y}{\sqrt{z}}+\frac{\sqrt{z}}{y}\right)\right]\,,
\end{equation}
which can be written as
\begin{equation}\label{final}
    Z_\qq\exp\Big[-|\beta_{j,\qq}|^2\left\{(N_\qq+1)(1-\ee^{\ri\omega_\qq
    t}) + N_\qq(1-\ee^{-\ri\omega_\qq
    t})\right\}\Big]\,,
\end{equation}
with $N_\qq=z/(1-z)=(\ee^{\hbar\omega_\qq/k_B T}-1)^{-1}$ and
$Z_\qq=1/(1-z)=(1-\ee^{-\hbar\omega_\qq/k_B T})^{-1}$. Taking the
impurity part and the product of all phonon modes $\qq$ into account
we find the complete memory function
\begin{eqnarray}\label{fullmemory}\fl\nonumber
    W_{i,j}(s)&=&2\delta_{j,i\pm 1}\Bigg(\frac{J}{\hbar}\Bigg)^2\mathrm{Re}\Bigg[\exp\Big\{\!-\sum_{\qq \neq 0}
    |\beta_{j,\qq}|^2 \\\fl
    &&\qquad\times\,[(N_\qq+1)(1-\ee^{\ri\omega_{\qq}s})+
    N_\qq(1-\ee^{-\ri\omega_{\qq}s})]\Big\}\exp(\pm\ri\omega_B
    t)\Bigg]\,,
\end{eqnarray}
where we used $Z_B = \prod_\qq Z_\qq$.

\section{Self-trapping}\label{aselftrap}

Impurities immersed in a BEC get self-trapped for sufficiently
strong impurity-BEC interactions, even in the absence of an
additional trapping potential
\cite{PhysRevB.46.301,cucchietti:210401,kalas:043608}. Based on
the results for static impurities in Section~\ref{static} we now
show that for the parameter regime considered in this paper
self-trapping effects can be neglected.

As pointed out by Gross~\cite{Gross-classical} the coupled
equations describing the impurity and the condensate in the
Hartree approximation are given by
\begin{eqnarray}\label{e1}
  \mu_b\,\psi(\rr)&=&-\frac{\hbar^2}{2m_b}\nabla^2\psi(\rr) + g|\psi(\rr)|^2\psi(\rr) +
  \kappa|\chi(\rr)|^2\psi(\rr)
  \\\label{e2}
  \varepsilon\chi(\rr)&=&-\frac{\hbar^2}{2m_a}\nabla^2\chi(\rr) +
  \kappa|\psi(\rr)|^2\chi(\rr)\,,
\end{eqnarray}
where $\chi(\rr)$ is the wavefunction of the impurity, $m_a$ is
the impurity mass and $\varepsilon$ the impurity energy. To
determine whether the impurity localizes for given experimental
parameters one has, in principle, to solve the coupled equations
for $\psi(\rr)$ and $\chi(\rr)$~\cite{kalas:043608}.
Alternatively, as suggested in~\cite{cucchietti:210401}, we use
the Gaussian mode-function $\chi_{j,\sigma}(\rr)$ in
Eq.~(\ref{chiapp}) as a variational wavefunction for the impurity,
with the spread $\sigma$ as a free parameter, and minimize the
total energy of the system in the regime $\alpha\ll 1$, where the
linearization of the GP equation is valid. For a homogenous
condensate, the potential energy of the impurity and the BEC is
$(-\cla{E}_p)$, and thus adding the kinetic energy of the impurity
yields the total energy
\begin{equation}\label{eimpp}
    \cla{E}(\sigma)= \cla{E}_\mathrm{kin}(\sigma) - \cla{E}_p(\sigma) =
    \frac{D}{4}\frac{\hbar^2}{m_a\sigma^2} -
    \frac{\kappa^2}{g\xi^D}G(0,\sigma)\,.
\end{equation}
The impurity localizes if $\cla{E}(\sigma)$ has a minimum for a
finite value of $\sigma$, which depends on the dimensionless
quantity
\begin{equation}\label{param}
   \alpha^\prime = \frac{|\kappa|}{g}\frac{m_a}{m_b}\,\alpha\sim\frac{\cla{E}_p}{\cla{E}_\mathrm{kin}}\,.
\end{equation}
In one dimension, we find by asymptotically expanding
$\cla{E}(\sigma)$ in the limit $\sigma/\xi\gg 1$ that there exists a
self-trapping solution for arbitrarily small $\alpha^\prime$ and
that
\begin{equation}\label{sigma1d}
    \sigma_{1D} =
    \sqrt{2\pi}\,\frac{\xi}{\alpha^\prime}\,.
\end{equation}
For the two-dimensional case, asymptotically expanding
$\cla{E}(\sigma)$ in the limit $\sigma/\xi\gg 1$ yields a critical
value $\alpha^\prime_c=2\pi$, above which self-trapping occurs.
The corresponding spread of the self-trapping solution
$\sigma_{2D}$ diverges close to $\alpha^\prime_c$, which validates
the asymptotic expansion of $\cla{E}(\sigma)$. In three
dimensions, numerical minimization of $\cla{E}(\sigma)$ shows that
the critical value is $\alpha^\prime_c \approx 31.7$, and the
corresponding self-trapped state is highly localized with
$\sigma_{3D} \approx 0.87\,\xi$.

According to Eq.~(\ref{sigma1d}) the spread of the self-trapping
solution exceeds several lattice spacings for the parameter regime
considered in this paper. In other words, the spread $\sigma_{1D}$
is much larger than the harmonic oscillator length $\sigma_\ell$
in practice, and hence self-trapping effects are indeed small.


\newpage

\section*{References}

\bibliographystyle{iopart-num}
\bibliography{njpref}

\end{document}